\documentclass[num-refs,sort&compress]{wiley-article}
\usepackage{float}
\usepackage{comment}
\usepackage{siunitx}
\usepackage{amsmath}
\usepackage[hidelinks]{hyperref}

\usepackage[framed]{matlab-prettifier}
\lstMakeShortInline"
\papertype{Preprint}

\title{Modeling animal contests based on spatio-temporal dynamics}

\author[1]{Amir Haluts$^\star$}
\author[2]{Alex Jordan}
\author[1]{Nir S. Gov}

\affil[1]{Department of Chemical and Biological Physics, Weizmann Institute of Science, Rehovot 7610001, Israel}
\affil[2]{Department of Collective Behavior, Max Planck Institute of Animal Behavior, Konstanz 78315,
Germany}

\corraddress{$^\star$Amir Haluts, Department of Chemical and Biological Physics, Weizmann Institute of Science, Rehovot 7610001, Israel}
\corremail{halutsamir@gmail.com}

\fundinginfo{
This work was supported by a Minna-James-Heineman-Stiftung
research grant (to Alex Jordan).}

\authorcontrib{A.H., A.J., and N.S.G. conceived and designed the study. A.H. and N.S.G performed modeling work. A.H. performed
simulations and analyzed output data. All authors wrote the paper.}

\runningauthor{Haluts et al. 2022. Modeling animal contests based on spatio-temporal dynamics}

\begin{document}
\begin{titlepage}
\pagenumbering{gobble}
\end{titlepage}
\cleardoublepage 
\pagenumbering{arabic}
\setlength{\parskip}{0pt}
\begin{frontmatter}
\maketitle

\begin{abstract}
\textbf{\uppercase{Abstract.}}
We present a general theoretical model for the spatio-temporal dynamics of animal contests. Inspired by interactions between physical particles, the model is formulated in terms of effective interaction potentials, which map typical elements of contest behaviour into empirically verifiable rules of contestant motion. This allows us to simulate the observable dynamics of contests in various realistic scenarios, notably in dyadic contests over a localized resource. Assessment strategies previously formulated in game-theoretic models, as well as the effects of fighting costs, can be described as variations in our model's parameters. Furthermore, the trends of contest duration associated with these assessment strategies can be derived and understood within the model. Detailed description of the contestants' motion enables the exploration of spatio-temporal properties of asymmetric contests, such as the emergence of chase dynamics. Overall, our framework aims to bridge the growing gap between empirical capabilities and theory in this widespread aspect of animal behaviour.

\keywords{\emph{animal contests}, \emph{agonistic interactions}, \emph{assessment strategy}, \emph{resource-holding potential}, \emph{tracking}, \emph{measuring animal behaviour}, \emph{trajectories}, \emph{effective interaction forces}, \emph{physics-inspired modeling}}
\end{abstract}
\end{frontmatter}

\setlength{\topskip}{0pt}
\section*{\uppercase{Introduction}}
Contests over limited resources are a common feature of animal behaviour, and have been the focus of many empirical and theoretical works over the past decades \cite{briffa2013introduction}. Due to the cost and benefit trade-offs that they entail, and following the foundations laid by the seminal works of the 1970s \cite{smith1973logic,smith1974theory,parker1974assessment,smith1976logic,parker1979sexual}, animal contests were predominantly modelled within the framework of game theory \cite{enquist1983evolution,payne1997animals,payne1998gradually,kokko2013dyadic}. Although they can generate empirically testable predictions---notably trends of contest duration and escalation \cite{arnott2009assessment}, experimental verification of game-theoretic contest models remains elusive \cite{taylor2003mismeasure,arnott2009assessment,pinto2019all,chapin2019further}, as empirical studies rarely yield more than anecdotal evidence in support of a particular class of models or the rejection of another \cite{arnott2009assessment,pinto2019all,chapin2019further}. Moreover, the theoretical foundations of these models are typically stated in terms of how contestants gather information about their own and their rival's 'resource holding potential' (RHP) \cite{payne1996escalation,mesterton1996wars,enquist1983evolution,enquist1990test,parker1981role,hammerstein1982asymmetric,payne1997animals,payne1998gradually,kokko2013dyadic}, a generally-defined measure for the ability to obtain and defend resources \cite{smith1976logic} that can rarely be measured directly. These difficulties have sparked disagreement over best practice in measuring animal contests \cite{taylor2003mismeasure,chapin2019further,parker2019so,elwood2019problems,leimar2019game}, and highlight the fact that most contest models meet the observable dynamics of contest behaviour only in their endpoint predictions---and ignore the detailed dynamics of contests in real time and space, making direct comparison of theoretical contest games with empirical observations of real animal contests inherently difficult. 

The most striking aspect of animal contests is the spatial dynamics of contestants as they react to real-time inputs. Importantly, the spatial dynamics of contests are directly measurable, and nowadays can be readily tracked \cite{dell2014automated,nathan2022big}. Although seemingly intricate and diverse \cite{Huntingford1987}, these dynamics commonly involve stereotypical behavioural elements that characterize contests in many species \cite{briffa2013introduction,ekman2009darwin,Huntingford1987}. Another central element is a spatially localized resource, commonly a mate \cite{jordan2014reproductive,emlen1977ecology}, or territory \cite{fuxjager2017animals}, which attracts potential rivals and drives them into contest range. Once the contestants are engaged in an interaction, their spatial dynamics are governed by behavioural elements that typify agonistic encounters, such as displays, attacks, and retreats \cite{Huntingford1987}. These features of animal contests can be described as universal rules of contestant motion, which, we propose, can then be associated with effective interaction forces that encode the contestants' behaviour, as we have recently shown for a system of spider contestants \cite{haluts2021spatiotemporal}. This approach yielded new mechanistic explanations for previous observations regarding the competitive advantage of larger contestants \cite{haluts2021spatiotemporal}. Similar methods have been applied to model inter-agent interactions in animal groups in the context of collective behaviour \cite{kovalev2021numerical,gorbonos2016long,hemelrijk2011some,hemelrijk2004density,huth1992simulation,bastien2020model}. This motivates us to propose a new theoretical framework for the observable dynamics of animal contests, which relies on generic and broadly applicable rules of inter-contestant interactions.

In this work, we construct a general theoretical model for the spatial and temporal dynamics of animal contests. The model is formulated in terms of effective interaction potentials, which map typical elements of contest behaviour into rules of attraction and repulsion between contestants, and are analogous to the potential interaction energies between physical particles. Through the effective interaction forces that they generate, these potentials govern the motion of contestants as they interact with each other and with a localized resource. This fundamental framework is used to simulate the spatio-temporal dynamics of our model's contestants in dyadic contests. Using simulated data, we demonstrate how our model's interaction potentials can be measured empirically in any system in which contest dynamics can be observed.

The scope of this general framework goes far beyond that of our motivating special case \cite{haluts2021spatiotemporal}. We show that the previously proposed RHP-assessment strategies \cite{payne1996escalation,mesterton1996wars,enquist1983evolution,enquist1990test,parker1981role,hammerstein1982asymmetric,payne1997animals,payne1998gradually,kokko2013dyadic}, which are stated in terms of how contestants gather information about their own and their rival's RHP, can be described as variations in the model's parameters. This is done by introducing an 'assessment function', which can describe various modes of assessment within its continuous parameter space. Further extending the relation between the model's parameters and the underlying behaviour, we account for fighting costs. Using the model's representation of the well-studied self- and mutual assessment strategies \cite{arnott2009assessment}, we show that the RHP-dependent trends of contest duration associated with these assessment strategies can be derived and understood within our model as an emergent property of contest dynamics. Finally, we explore spatio-temporal properties of asymmetric contests between RHP-unmatched contestants. These results showcase the applicability of the model to various realistic contest scenarios, in which the comparison between theory and experiment can be facilitated by the analysis of spatio-temporal data derived from the contestants' trajectories.

\section*{\uppercase{The basic model: construction and measurement}}
\subsection*{Effective Interaction Potentials}
Our model is based on the mapping of typical contest behaviour to generic rules of inter-contestant interactions. These rules are encoded by effective 'contestant interaction potentials' $V_{j\rightarrow i}$, which capture the influence of a rival contestant $j$ on the motion of contestant $i$
(Fig. \ref{fig1}\textit{A},\textit{B}). Not to be confused with resource holding potentials (RHPs), our interaction potentials are analogous to the potential interaction energies that govern the interactions between physical particles. We construct these potentials based on the following generic features of contest behaviour, which we state in terms of effective attraction or repulsion between $i$ and $j$ depending on the distance between them: (1) Long-range repulsion due to mutual avoidance (which can be surmounted due to an attracting resource, as shown in Fig. \ref{fig1}\textit{C},\textit{D}), (2) Medium- to short-range attraction when the contestants reach a separation distance in which conflict escalation is inevitable (and hence move towards each other), (3) Strong repulsion at contact, and (4) The strength of the interaction decays to zero when the contestants are far apart. Note that the tendency to decrease the inter-contestant distance (effective attraction) is associated here with conflict escalation, while the tendency to increase this distance (effective repulsion) is associated with de-escalation. These effects can be directly measured in experiments, as we demonstrate in a later section.

Various interaction potentials can be constructed to satisfy the above requirements, that is to have a qualitative shape as in Fig. \ref{fig1}\textit{A},\textit{B}. Here we propose one such particular potential, a combination of a logarithmic repulsion and an attractive Gaussian well, for which the extrema can be obtained analytically (Supporting Information S1), 
\begin{equation}\label{eq1}
    V_{j\rightarrow i}(\mathrm{x}_{ij}) = - \alpha_{j\rightarrow i}\exp(-\beta{\mathrm{x}_{ij}}^2) - \delta_{j\rightarrow i}\ln(\mathrm{x}_{ij}),\qquad\mathrm{x}_{ij} = \frac{|\mathbf{r}_i - \mathbf{r}_j|}{\mathrm{x}_0}
\end{equation}
where $\mathrm{x}_{ij}$ is the (dimensionless) distance between contestants $i$ and $j$ (with respective position vectors $\mathbf{r}_i$ and $\mathbf{r}_j$), $\alpha_{j\rightarrow i}$ and $\delta_{j\rightarrow i}$ are positive 'interaction parameters' that set the strengths of effective attraction and repulsion (experienced by $i$ when interacting with a rival $j$), $\beta > 0$ determines the range of effective attraction, and $\mathrm{x}_0$ is a length scale distance parameter. Both $\beta$ and $\mathrm{x}_0$ are assumed to be the same for all contestants in a given system, and in this work will be set to equal $1$. Supporting Information S2 addresses the addition of an intermediate 'evaluation' regime, which has been observed in various animal contests \cite{haluts2021spatiotemporal,enquist1983evolution,enquist1990test,jensen1998aggression}, to $V_{j\rightarrow i}$, demonstrating the inclusion of other (system-specific) features in this interaction potential.

Note that $\alpha_{j\rightarrow i}$ reflects the motivation of contestant $i$ to escalate the interaction, and is therefore associated with the (absolute or relative) RHP of $i$, while $\delta_{j\rightarrow i}$ reflects how intimidating (repulsive) the rival $j$ is perceived by $i$, and is therefore associated with the (absolute or relative) RHP of $j$. In a later section, we construct an explicit functional relationship between the parameters $\alpha_{j\rightarrow i}$ and $\delta_{j\rightarrow i}$ and the contestants' RHPs depending on the assessment strategy that they employ. An analogous interaction potential $V_{i\rightarrow j}$ (with interaction parameters $\alpha_{i\rightarrow j}$ and $\delta_{i\rightarrow j}$) is experienced by the rival $j$ due to the interaction with $i$ and, as implied by the directional arrow notation, in general $V_{j\rightarrow i} \ne V_{i\rightarrow j}$ since these potentials manifest asymmetries between the contestants (as illustrated in Fig. \ref{fig1}\textit{E}). The influence of $V_{j\rightarrow i}$ and $V_{i\rightarrow j}$ on the motion of the contestants is governed by the non-reciprocal forces that they generate, $F_{j\rightarrow i} = -d V_{j\rightarrow i}/d\mathrm{x}_{ij}$ and $F_{i\rightarrow j} = -d V_{i\rightarrow j}/d\mathrm{x}_{ij}$.

Access to limited resources is the primary motivation of animals to engage in contests in the first place \cite{briffa2013introduction}. Most of these resources, notably mates, food, and territories, are inherently spatially localized, and therefore act as attracting regions in space that drive competitors closer to each other until conflict is inevitable. The level of attraction towards a given resource is determined by the perceived resource value, which could vary not only according to an absolute scale, but also between contestants and contexts \cite{kokko2013dyadic}. We model the influence of a localized resource on the motion of contestant $i$ as an effective 'resource potential' $V_{\mathrm{res}\rightarrow i}$. As with $V_{j\rightarrow i}$, $V_{\mathrm{res}\rightarrow i}$ should not be confused with the concept of RHP, but rather should be thought of as the effective 'energy' landscape created by the resource. The particular shape of the resource potential's landscape may be system-specific \cite{haluts2021spatiotemporal}, but its global qualitative effect is generic: to bring contestants into contest range due to their mutual attraction to the resource. A resource potential with a simple radially-symmetrical form is shown in Fig. \ref{fig1}\textit{C},\textit{D}. Note that the effective potential landscapes of Fig. \ref{fig1}\textit{A} and \textit{C} describe interactions in a two-dimensional (planar) space, but the model is equally applicable in three dimensions.

Once contestants become strongly engaged in a contest interaction, their attention is predominantly given to their rival until the encounter is resolved. This implies an 'attention switch' in the interactions with a resource and with rivals, where the contestants' motion is significantly affected by their attraction to the resource (i.e. by $V_{\mathrm{res}\rightarrow i}$ and $V_{\mathrm{res}\rightarrow j}$) only when they are relatively far apart, and is dominated by their interaction with each other (i.e. by $V_{j\rightarrow i}$ and $V_{i\rightarrow j}$) when they are within the contest range. This attention switch can be expressed by the total effective potential experienced by contestant $i$, $V_{\mathrm{tot}\rightarrow i}$, which combines the influence of a rival $j$ and of a resource on the motion of $i$,
\begin{equation}\label{eq2}
V_{\mathrm{tot}\rightarrow i}(\mathbf{r}_i,\mathbf{r}_j) =
\begin{cases}
    V_{\mathrm{res}\rightarrow i}(\mathbf{r}_i) + V_{j\rightarrow i}(\mathrm{x}_{ij}), & \mathrm{x}_{ij} \geq \mathrm{x}_\cap\qquad\text{(attention to resource + rival)}\\
    V_{j\rightarrow i}(\mathrm{x}_{ij}), & \mathrm{x}_{ij} < \mathrm{x}_\cap\qquad\text{(attention to rival)}
\end{cases}
\end{equation}
where $\mathrm{x}_\cap$ is the contest onset distance, as defined below and in Fig. \ref{fig1}\textit{F}. A respective total effective potential $V_{\mathrm{tot}\rightarrow j}$ is experienced by the rival $j$ due to $i$ and the resource. Below we demonstrate how these potentials can be extracted from contest trajectories, and in ref. \cite{haluts2021spatiotemporal} we demonstrate their extraction in a specific system of spider contestants, where the resource potential has a non-trivial shape.

\subsection*{Definition of a Contest}
\begin{figure}[h!]
    \centering
    \includegraphics[width = 14.1cm]{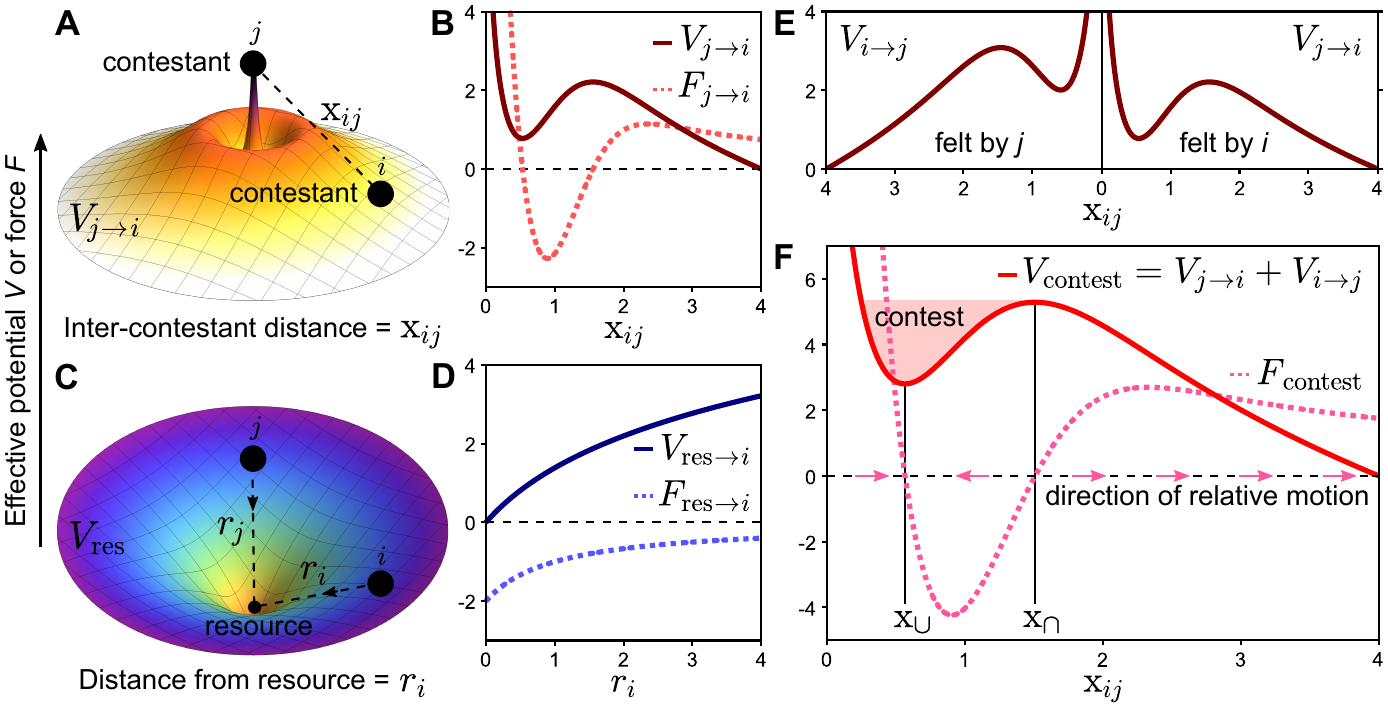}
    \caption{\textbf{Effective interaction potentials.} (\textbf{\textit{A}}) The landscape of the effective 'contestant interaction potential' $V_{j\rightarrow i}$, as given in Eq. \eqref{eq1}, with interaction parameters $\alpha_{j\rightarrow i} = 7$, $\delta_{j\rightarrow i} = 3$, and $\beta = 1$. (\textbf{\textit{B}}) The profiles of $V_{j\rightarrow i}$ and its corresponding force $F_{j\rightarrow i} = -d V_{j\rightarrow i}/d\mathrm{x}_{ij}$ as a function of the inter-contestant distance $\mathrm{x}_{ij}$. The graph of $V_{j\rightarrow i}$ was shifted vertically such that the lowest shown point has a 'height' of zero. Parameter values as in \textit{A}. (\textbf{\textit{C}}) The landscape of an effective 'resource potential' $V_\mathrm{res}$ with a simple radially-symmetrical form. Resources act as attracting regions in space that drive contestant into conflict range. (\textbf{\textit{D}}) The profiles of $V_{\mathrm{res}\rightarrow i}$ and its corresponding force $F_{\mathrm{res}\rightarrow i} = -d V_{\mathrm{res}\rightarrow i}/d r_i$ as a function of the distance from the resource $r_i = |\mathbf{r}_i|$. \textit{C} and \textit{D} show the potential $V_{\mathrm{res}\rightarrow i}(\mathbf{r}_i) = p_i\ln(r_i + \epsilon)$, where $p_i > 0$ sets the attractiveness of the resource (as perceived by contestant $i$), and $\epsilon > 0$ prevents divergence at the resource (defined as the origin), with $p_i = 4$ and $\epsilon = 1$. (\textbf{\textit{E}}) Asymmetries between contestants are manifested by their contestant interaction potentials (in general $V_{j\rightarrow i} \ne V_{i\rightarrow j}$). Here $V_{j\rightarrow i}$ is the same as in \textit{B}, and $V_{i\rightarrow j}$ is shown with $\alpha_{i\rightarrow j} = 8$, $\delta_{i\rightarrow j} = 4$, and $\beta = 1$. (\textbf{\textit{F}}) The relative 'contest force' $F_\mathrm{contest} = F_{j\rightarrow i} + F_{i\rightarrow j}$ governs the relative motion along the inter-contestant axis (arrows). This force is associated with a relative 'contest potential' $V_\mathrm{contest} = V_{j\rightarrow i} + V_{i\rightarrow j}$, which has its local maximum ($\mathrm{x}_\cap$) and minimum ($\mathrm{x}_\cup$) where $F_\mathrm{contest} = 0$ (no relative motion between the contestants). The distance $\mathrm{x}_{ij} = \mathrm{x}_\cap$ is defined as the contest onset.}
    \label{fig1}
\end{figure}

In order to clearly define the onset of a 'contest' in our model, we consider the relative motion between the contestants along the inter-contestant axis (Fig. \ref{fig1}\textit{A}) due only to $V_{j\rightarrow i}$ and $V_{i\rightarrow j}$. Note that according to Eq. \eqref{eq1}, this relative motion depends only on the distance between the contestants, and is governed by the effective forces $F_{j\rightarrow i}$ and $F_{i\rightarrow j}$. Taking the $j\rightarrow i$ direction as the 'positive' direction $\hat{\mathbf{x}}_{ij} = (\mathbf{r}_i - \mathbf{r}_j)/|\mathbf{r}_i - \mathbf{r}_j|$, and the position of contestant $j$ as a fixed point of reference, contestant $i$ appears to be driven along $\hat{\mathbf{x}}_{ij}$ by a relative 'contest force' $F_\mathrm{contest}$, given by
\begin{equation}\label{eq3}
    F_\mathrm{contest}(\mathrm{x}_{ij}) = F_{j\rightarrow i}(\mathrm{x}_{ij}) + F_{i\rightarrow j}(\mathrm{x}_{ij}) = -\frac{d}{d \mathrm{x}_{ij}}\left[V_{j\rightarrow i}(\mathrm{x}_{ij}) + V_{i\rightarrow j}(\mathrm{x}_{ij})\right]
\end{equation}
where note that the sum is taken since $F_{i\rightarrow j}$, which is applied by $i$ on $j$ in the $-\hat{\mathbf{x}}_{ij}$ direction, appears to drive $i$ relative to $j$ in the opposing $+\hat{\mathbf{x}}_{ij}$ direction. Eq. \eqref{eq3} motivates the definition of a relative 'contest potential' $V_\mathrm{contest}$ as the sum of the individual interaction potentials,
\begin{equation}\label{eq4}
    V_\mathrm{contest}(\mathrm{x}_{ij}) = V_{j\rightarrow i}(\mathrm{x}_{ij}) + V_{i\rightarrow j}(\mathrm{x}_{ij})
\end{equation}
such that $F_\mathrm{contest} = -d V_\mathrm{contest}/d\mathrm{x}_{ij}$. We use Eqs. \eqref{eq3} and \eqref{eq4} to define the onset of a contest according to the direction of relative motion, which changes at the locations of the local maximum ($\mathrm{x}_\cap$) and minimum ($\mathrm{x}_\cup$) of $V_\mathrm{contest}$ (see Supporting Information S1 for the analytic expressions of these extrema). When the contestants reach a separation $\mathrm{x}_{ij}$ that is shorter than $\mathrm{x}_\cap$, the relative force becomes attractive ($F_\mathrm{contest} < 0$) towards $\mathrm{x}_\cup$, and the interaction is governed by a transient bounded state (see Fig. \ref{fig1}\textit{F}). This bounded state is equivalent to the ultimate escalation into a short-range contest, in which the contestants are completely engaged with each other. We therefore define $\mathrm{x}_\cap$ as the contest onset distance, and regard two contestants as engaged in a contest when $\mathrm{x}_{ij} < \mathrm{x}_\cap$ (Fig. \ref{fig1}\textit{F}). Note that according to the particular choice of Eq. \eqref{eq1}, the relative contest potential is given by
\begin{equation}\label{eq5}
     V_\mathrm{contest}(\mathrm{x}_{ij}) = - (\alpha_{j\rightarrow i} + \alpha_{i\rightarrow j})\exp(-\beta{\mathrm{x}_{ij}}^2) - (\delta_{j\rightarrow i} + \delta_{i\rightarrow j})\ln(\mathrm{x}_{ij}).
\end{equation}

To simulate the dynamics of our model's contestants, we treat them as Brownian particles moving in space under the influence of an external potential \cite{romanczuk2012active}, as described in Supporting Information S3. Note that in this work, all contests were simulated in a two-dimensional (planar) space. Fig. \ref{fig2}\textit{A}--\textit{C} describe typical trajectories of a simulated symmetric interaction between two identical contestants ($V_{j\rightarrow i} = V_{i\rightarrow j}$), where the attraction of both contestants to the resource brings them into contest range.

\vspace{-10pt}

\subsection*{Extracting Interaction Potentials from Empirical Measurements}
\begin{figure}[h!]
    \centering
    \includegraphics[width = 14.1cm]{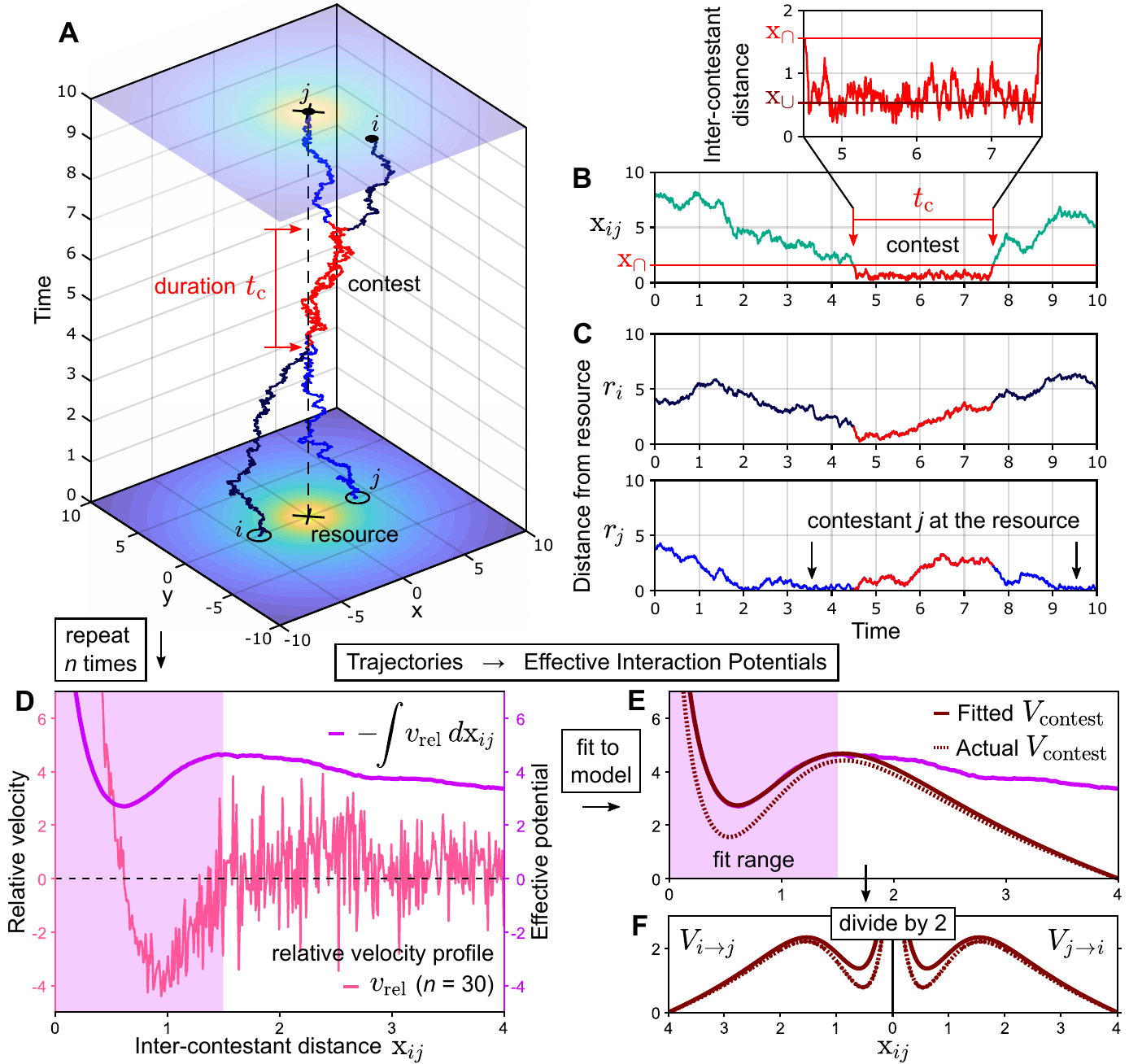}
    \caption{(\textit{\textbf{A}}--\textit{\textbf{C}}) \textbf{Contestant dynamics.} (\textit{\textbf{A}}) Typical simulated trajectories of two identical contestants ($V_{j\rightarrow i} = V_{i\rightarrow j}$) in the vicinity of a resource. Each contestant's dynamics was simulated using Eq. \ref{eqS9} with $\eta = 1$ and $D = 0.5$ (see Supporting Information S3), and $V_{\mathrm{tot}\rightarrow i}$ as defined in Eq. \eqref{eq2}, with $V_{j\rightarrow i}$ and $V_\mathrm{res}$ as in Fig. \ref{fig1}\textit{A} and \textit{C}, respectively. The contestants were initialized at equivalent positions on opposing sides of the resource, with $\mathbf{r}_i(t = 0) = (-4, 0)$ and $\mathbf{r}_j(t = 0) = (4, 0)$, as marked by the empty circles in the x--y plane. Segments of the trajectories in which the contestants were engaged in a contest ($\mathrm{x}_{ij} < \mathrm{x}_\cap$) are shown in red. (\textit{\textbf{B}}) The distance between the contestants throughout the simulation. The tendency of the contestants to get closer to each other due to their mutual attraction to the resource brings them into contest range. As evident in the close-up view, the contestants spend the majority of the contest near the minimum, $\mathrm{x}_\cup$, of the contest potential $V_\mathrm{contest}$. (\textit{\textbf{C}}) The distance between each contestant and the resource (the origin) throughout the simulation. In this particular instance, contestant $j$ reached the resource before the contest, and regained it after the contest. (\textit{\textbf{D}}--\textit{\textbf{F}}) \textbf{Extracting interaction potentials from trajectories.} (\textit{\textbf{D}}) The averaged relative velocity profile $v_\mathrm{rel}$ and the observed relative potential $V_\mathrm{rel}$ obtained by integration, as calculated from the trajectories of $n = 30$ simulated contests as in \textit{A}. The shaded rectangle marks the fit range used in \textit{E}. (\textit{\textbf{E}}) Fit of the model's contest potential to $V_\mathrm{rel}$ within the range dominated by $V_\mathrm{contest}$. This fit deviates only slightly from the actual $V_\mathrm{contest}$. (\textit{\textbf{F}}) For these symmetric contests, the fitted $V_\mathrm{contest}$ is simply divided by 2 to obtain $V_{j\rightarrow i}$ and $V_{i\rightarrow j}$. See also supporting information S4.
    }
    \label{fig2}
\end{figure}

Using simulated contestant trajectories as in Fig. \ref{fig2}\textit{A}, for which the underlying potentials are known, we demonstrate how the contestants' effective interaction potentials can be extracted from the spatio-temporal dynamics of contests. These potentials govern the velocity at which the contestants tend to move with respect to one another depending on their position. With minimal \textit{a priori} assumptions about the observed interactions, this relative motion can be gauged by the averaged relative velocity between the contestants as a function of $\mathrm{x}_{ij}$,
\begin{equation} \label{eq6}
    v_\mathrm{rel}(\mathrm{x}_{ij}) = \text{mean}\left(\frac{\Delta\mathrm{x}_{ij}}{\Delta t}(\mathrm{x}_{ij})\right),
\end{equation}

where note that $v_\mathrm{rel} > 0$ (a tendency to increase $\mathrm{x}_{ij}$) indicates effective repulsion between the contestants, while $v_\mathrm{rel} < 0$ (a tendency to decrease $\mathrm{x}_{ij}$) indicates effective attraction. Fig. \ref{fig2}\textit{D} shows the relative velocity profile described by $v_\mathrm{rel}$, as calculated from the trajectories of $n = 30$ simulated contests as in Fig. \ref{fig2}\textit{A}. To be informative, $v_\mathrm{rel}$ requires sufficient sampling, such that the stochastic components of the motion are mostly averaged out, and that the sampled contests are comparable, so that the remaining averaged dynamics represents the data well. For example, the relative velocity profile of Fig. \ref{fig2}\textit{D} represents a fairly large but experimentally-feasible set of RHP-matched contests with the same initial setup (comparable contests).

Then, an averaged relative interaction potential can be derived from $v_\mathrm{rel}$ by integration with respect to $\mathrm{x}_{ij}$,
\begin{equation} \label{eq7}
    \eta V_\mathrm{rel}(\mathrm{x}_{ij}) \approx -\int_0^{\mathrm{x}_{ij}} v_\mathrm{rel}(\mathrm{x}^\prime_{ij})\,d\mathrm{x}^\prime_{ij}
\end{equation}
where the 'integral' denotes the approximate cumulative integration over the discrete values of the measured $v_\mathrm{rel}$ (see supporting information S4 for the formal definition). Eq. \eqref{eq7} assumes that the effects of the contestants' inertia can be neglected \textit{on average}, such that the averaged relative velocity is directly proportional (through the 'mobility' $\eta$, see Supporting Information S3) to the effective relative force associated with $V_\mathrm{rel}$. In practice, $\eta$ can be simply absorbed into $V_\mathrm{rel}$ by setting $\eta = 1$. Fig. \ref{fig2}\textit{D} shows the result of this integration, and Supporting Information S4 describes the full practical implementation of Eqs. \eqref{eq6} and \eqref{eq7} to a set of contest trajectories.

Although the overall shape of the observed potential $V_\mathrm{rel}$ is in good qualitative agreement with the actual contest potential, it includes contributions from the residual attraction of both contestants towards the resource, and therefore cannot be strictly identified with $V_\mathrm{contest}$. By observing the typical contest behaviour of the studied animals, one should be able to estimate the average inter-contestant distance at which the 'attention switch' of Eq. \eqref{eq2} comes into play, and thereby estimate the range of $\mathrm{x}_{ij}$ within which the interaction is dominated by $V_\mathrm{contest}$ and the effect of the resource is negligible. For example, a clear change of the averaged relative motion from effective repulsion to effective attraction, as indicated by a sign change of $v_\mathrm{rel}$, is a possible indicator for the onset of this range. In this range, a model for $V_\mathrm{contest}$ can be fitted to $V_\mathrm{rel}$ (supporting information S4), as illustrated in Fig. \ref{fig2}\textit{E}. With the $n = 30$ simulated trajectories used here, a fit according to Eq. \eqref{eq5} deviates only slightly from the actual $V_\mathrm{contest}$.

Finally, for contests between RHP-matched contestants as in the current example, the fitted $V_\mathrm{contest}$ can be simply divided by 2 to obtain the contestant interaction potentials $V_{j\rightarrow i}$ and $V_{i\rightarrow j}$, as shown in Fig. \ref{fig2}\textit{F}. Otherwise, a relation between the contestants' RHP and the parameters of the interaction potentials (an RHP-assessment strategy) has to be assumed in order to dissect $V_\mathrm{contest}$ into $V_{j\rightarrow i}$ and $V_{i\rightarrow j}$. To avoid such \textit{a priori} assumptions about the underlying assessment strategy, we propose that the initial extraction of effective interaction potentials should always rely on contests sampled from an RHP-uniform pool of contestants, e.g. contestants of similar sizes in systems where size is strongly correlated with RHP. Once effective interaction potentials are established for these RHP-matched contestants, evidence for the assessment strategy can be gathered, based on the theory developed in the following section, by sampling contests between contestants of other RHPs.

\section*{\uppercase{Integrating assessment and costs}}

\subsection*{The Assessment Function}
The type of RHP-related information used by contestants to resolve contests, commonly termed the 'RHP-assessment strategy', has been widely used to classify contests in different species \cite{arnott2008information,arnott2009assessment}. Historically, theoretical and empirical studies have analyzed and classified animal contests according to two main categories of assessment: self-assessment, in which contestants only consider their own RHP and do not gather information about their rival's RHP \cite{arnott2009assessment}, and mutual assessment, in which contestants consider both their own and their rival's RHP in their decision-making \cite{arnott2009assessment}. Existing game-theoretic models of contest behaviour were mostly constructed according to a specific assessment strategy of either the self- or mutual assessment categories, and this often limits their applicability only to systems which closely follow their underlying assessment paradigm  \cite{chapin2019further}.

Here we propose that any mode of assessment can be expressed in terms of how the parameters $\alpha_{j\rightarrow i}$ and $\delta_{j\rightarrow i}$ of Eq. \eqref{eq1} vary with the contestants' RHP. We assume that the RHP of contestants $i$ and $j$ can be sufficiently expressed by single numeric variables, which are respectively denoted $m_i$ and $m_j$. Henceforth, we will often refer to $m_i$ and $m_j$ as the contestants' (effective) sizes, although in general RHP is not strictly interchangeable with size \cite{smith1976logic}. Nevertheless, any metric or proxy of RHP that can be numerically expressed is compatible with this approach. In order to facilitate comparisons across taxa and scales, it is useful to define
\begin{equation}\label{eq8}
    m_i = \frac{\text{RHP of contestant $i$}}{\text{RHP of reference}},
\end{equation}
such that $m_i$ and $m_j$ are dimensionless effective 'sizes' that express the RHP of the contestants relative to some chosen reference in the population of interest, e.g. the estimated average.

We construct an 'assessment function' $\mathbf{A}_{j\rightarrow i}$ (of contestant $i$ with respect to a rival $j$) which can express, in principle, any assessment strategy. For a two-component interaction potential as in Eq. \eqref{eq1}, $\mathbf{A}_{j\rightarrow i}$ can be represented as a two-element vector that defines the functional relationship between the parameters $\alpha_{j\rightarrow i}$ and $\delta_{j\rightarrow i}$ and the effective size variables $m_i$ and $m_j$. As explained above, $\alpha_{j\rightarrow i}$ is directly associated with the (absolute or relative) size of contestant $i$, while $\delta_{j\rightarrow i}$ is directly associated with the (absolute or relative) size of the rival $j$, and these features can be expressed by $\mathbf{A}_{j\rightarrow i}$ as follows,
\begin{equation}\label{eq9}
    \mathbf{A}_{j\rightarrow i}(m_i, m_j) =
    \begin{pmatrix}
    \alpha_{j\rightarrow i} \\ 
    \delta_{j\rightarrow i} 
    \end{pmatrix} = 
    \begin{pmatrix}
    \alpha_0\,{m_i}^s(m_i/m_j)^{s_Q} \\ 
    \delta_0\,{m_j}^r(m_j/m_i)^{r_Q} 
    \end{pmatrix}
\end{equation}
where $\alpha_0$ and $\delta_0$ are scaling parameters, and the different RHP-dependent components represent: ${m_i}^s$---absolute self-assessment; ${m_j}^r$---absolute rival-assessment; $(m_i/m_j)^{s_Q}$ and $(m_j/m_i)^{r_Q}$---relative self- and rival-assessment (mutual assessment), respectively. Note that the opponent's assessment function, $\mathbf{A}_{i\rightarrow j} = \left(\alpha_{i\rightarrow j},\delta_{i\rightarrow j}\right)$, is obtained by swapping the functional roles of $m_i$ and $m_j$ in Eq. \eqref{eq9}. The assessment strategy, which is now defined in terms of how $\alpha_{j\rightarrow i}$ and $\delta_{j\rightarrow i}$ vary with $m_i$ and $m_j$, is represented by the (non-negative) values of the exponents $s$, $r$, $s_Q$, $r_Q$. In particular, we propose that 'pure' self-assessment can be described by $s > 0$ and $r,\,s_Q,\,r_Q = 0$, that is
\begin{equation}\label{eq10}
    \mathbf{A}_{j\rightarrow i}(m_i, m_j) =
    \begin{pmatrix}
    \alpha_0\,{m_i}^s \\ 
    \delta_0 
    \end{pmatrix}\quad\text{for pure self-assessment},
\end{equation}
and that 'pure' mutual (relative) assessment can be described by $s,\,r = 0$ and $s_Q,\,r_Q > 0$, that is
\begin{equation}\label{eq11}
    \mathbf{A}_{j\rightarrow i}(m_i, m_j) =
    \begin{pmatrix}
    \alpha_0\,(m_i/m_j)^{s_Q} \\ 
    \delta_0\,(m_j/m_i)^{r_Q} 
    \end{pmatrix}\quad\text{for pure mutual assessment}.
\end{equation}
Importantly, any combination of these modes of assessment can be expressed by the assessment function $\mathbf{A}_{j\rightarrow i}$, and this allows the description of animal contests that do not follow a pure assessment strategy \cite{chapin2019further}.

\begin{figure}[h!]
    \centering
    \includegraphics[width = 14.1cm]{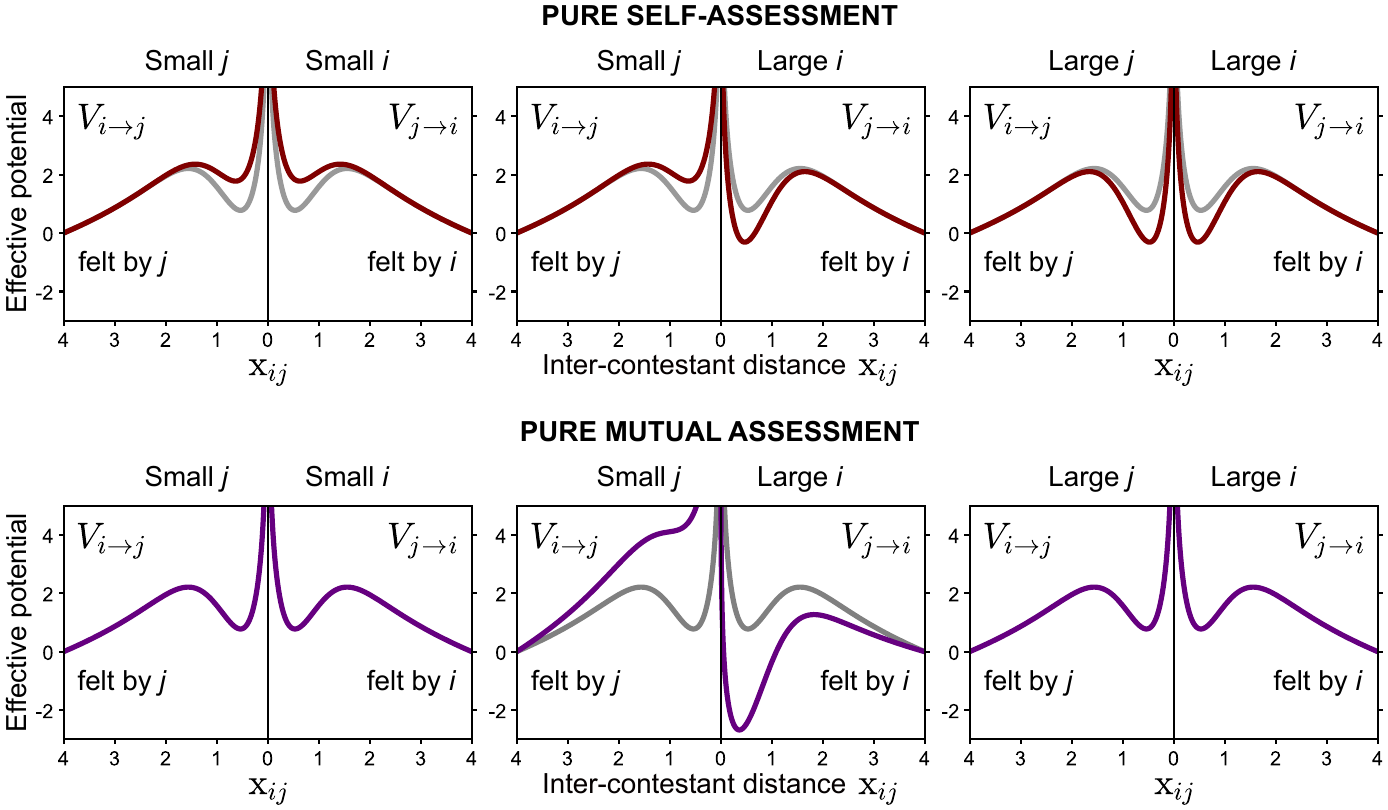}
    \caption{\textbf{Assessment strategies in terms of contestant interaction potentials.} By expressing an assessment strategy in terms of how the interaction parameters $\alpha_{j\rightarrow i}$ and $\delta_{j\rightarrow i}$ of Eq. \eqref{eq1} vary with the contestants' effective sizes, as described in Eqs. \eqref{eq8}--\eqref{eq13}, we map these behavioural categories into the interaction potentials of the contestants. Under pure self-assessment (Eq. \eqref{eq10} with $s = 1$) and pure mutual assessment (Eq. \eqref{eq11} with $s_Q = 1$ and $r_Q = 2$), the potentials $V_{i\rightarrow j}$ and $V_{j\rightarrow i}$ are shown for the interaction of small size-matched contestants ($m_j = m_i = 0.8$, \textit{Left}), a small contestant with a large contestant ($m_j = 0.8,\,m_i = 1.2$ for self-assessment, and $m_j = 0.9,\,m_i = 1.1$ for mutual assessment, \textit{Center}), and large size-matched contestants ($m_j = m_i = 1.2$, \textit{Right}). For reference, each graph also shows the potentials of medium size-matched contestants ($m_j = m_i = 1$) in grey. Eqs. \eqref{eq10} and \eqref{eq11} were used with $\alpha_0 = 7$ and $\delta_0 = 3$.}
    \label{fig3}
\end{figure}

Representing the attractive and repulsive components of Eq. \eqref{eq1} as a two-element vector $\mathbf{V}_0$,
\begin{equation}\label{eq12}
    \mathbf{V}_0(\mathrm{x}_{ij}) = 
    -\begin{pmatrix} 
    \exp(-\beta\mathrm{x}_{ij}^2) \\ 
    \ln(\mathrm{x}_{ij}) 
    \end{pmatrix}
\end{equation}
we can conveniently express $V_{j\rightarrow i}$ as the dot product of Eqs. \eqref{eq9} and \eqref{eq12},
\begin{equation}\label{eq13}
    V_{j\rightarrow i}(\mathrm{x}_{ij}) = \mathbf{A}_{j\rightarrow i}(m_i, m_j) \cdot \mathbf{V}_0(\mathrm{x}_{ij})
\end{equation}
This separation of contest behaviour into two functional components, a system-specific component that encodes the contestant's assessment strategy ($\mathbf{A}_{j\rightarrow i}$) and a more generic component that encodes the spatial properties of the contest interaction ($\mathbf{V}_0$), is an important feature of our model.

In Fig. \ref{fig3}, Eqs. \eqref{eq10}--\eqref{eq13} are used to plot the potentials $V_{i\rightarrow j}$ and $V_{j\rightarrow i}$ under pure self-assessment (Eq. \eqref{eq8} with $s = 1$) and pure mutual assessment (Eq. \eqref{eq9} with $s_Q = 1$ and $r_Q = 2$), for the interactions of small size-matched contestants, a small contestant with a large contestant, and large size-matched contestants. The differences between these assessment strategies are immediately apparent when comparing their respective interaction potentials. Notably, the self-assessment potentials vary with the absolute values of $m_i$ and $m_j$, while the mutual assessment potentials depend only on the size ratio $m_i/m_j$. This makes mutual assessment scale-invariant (the effective interaction potentials of size-matched contestants are independent of size, compare Fig. \ref{fig3} \textit{Left} and \textit{Right}), but very sensitive to size difference, as evident by the pronounced broken symmetry between $V_{j\rightarrow i}$ and $V_{i\rightarrow j}$ for the interaction of unmatched contestants (Fig. \ref{fig3} \textit{Center}).

\subsection*{Accounting for Fighting Costs}
The RHP is in general time-dependent, as it reflects the current state of a contestant. For example, injured or energy-depleted individuals would have a decreased RHP compared to their uncompromised state. Direct costs of fighting come in two main forms: (1) self-inflicted costs (due to a contestant's own actions), and (2) costs inflicted by the rival. The contributions of these costs to the contestants' decision-making have been previously considered as a feature of the assessment strategy \cite{arnott2009assessment}. In particular, pure self-assessment models assume that the actions of the rival do not inflict costs, while in mutual assessment both self- and rival-inflicted costs are integrated \cite{arnott2009assessment}. 

\begin{figure}[h!]
    \centering
    \includegraphics[width = 14.1cm]{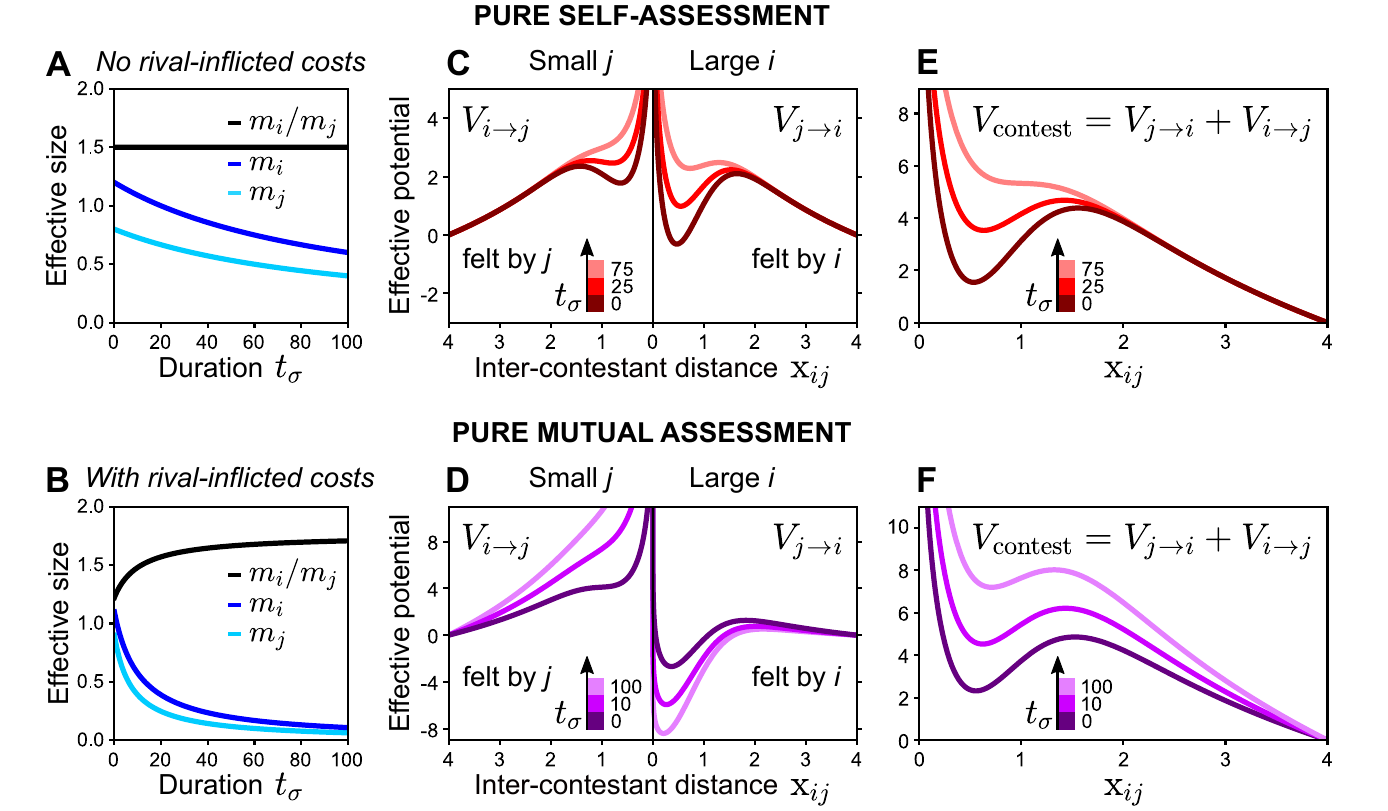}
    \caption{\textbf{Effects of fighting costs.} (\textit{\textbf{A}},\textit{\textbf{B}}) The dynamics of the effective sizes $m_i$, $m_j$, and of the ratio $m_i/m_j$, according to Eq. \eqref{eq16}. (\textit{\textbf{A}}) With only self-inflicted costs ($K_{ij} = 0$), where $K_\mathrm{self} = 0.01$, and with initial effective sizes $\mu_i = 1.2$ and $\mu_j = 0.8$. (\textit{\textbf{B}}) With both self-inflicted and rival-inflicted costs ($K_{ij} \ne 0$), where $K_\mathrm{self} = 0.01$ and $K_{ij} = 0.1$, and with initial effective sizes $\mu_i = 1.1$ and $\mu_j = 0.9$. (\textit{\textbf{C}},\textit{\textbf{D}}) The dependence of $m_i$ and $m_j$ on the accumulated contest duration $t_\sigma$ due to costs means that the interaction potentials themselves depend on $t_\sigma$. Here, the potentials $V_{i\rightarrow j}$ and $V_{j\rightarrow i}$ vary according to the values of $m_i$ and $m_j$ in \textit{A} (\textit{C}) and \textit{B} (\textit{D}) as $t_\sigma$ increases. Note that in \textit{D}, the asymmetry between $V_{i\rightarrow j}$ and $V_{j\rightarrow i}$ increases with $t_\sigma$, in accordance with the trend of $m_i/m_j$ in \textit{B}. (\textit{\textbf{E}},\textit{\textbf{F}}) The relative contest potential vary accordingly with $t_\sigma$, such that the bounding well that defines the contest regime becomes less attractive (shallower).
    }
    \label{fig4}
\end{figure}

Here we extend the model to account for these two main costs, but note that the following approach can be used, in principle, to account for any cost. We set the total self-inflicted cost and the total rival-inflicted cost as increasing functions of the accumulated contest time $t_\sigma$, and model the dependence of the effective size on these costs as
\begin{equation} \label{eq14}
    m_i(t_\sigma) = \frac{\mu_i}{1\,+\,C_i(t_\sigma)\,+\,C_{j\rightarrow i}(t_\sigma)},
\end{equation}
where $\mu_i$ is the effective size of contestant $i$ before the contest had started, $C_i$ is the accumulated self-inflicted cost, and $C_{j\rightarrow i}$ is the accumulated cost inflicted by the rival $j$. As a proof of concept, we write simple but useful expressions for $C_i$ and $C_{j\rightarrow i}$ as
\begin{equation} \label{eq15}
    C_i(t_\sigma) = K_\mathrm{self}\,t_\sigma \quad\text{and}\quad C_{j\rightarrow i}(t_\sigma) = K_{ij}\,\frac{\mu_j}{\mu_i}\,t_\sigma,
\end{equation}
where $K_\mathrm{self}$ and $K_{ij}\,\mu_j/\mu_i$ are the rates at which their respective costs are accrued. Eq. \eqref{eq15} naively assumes that costs are incurred continuously and deterministically, and that the rate of incurring self-inflicted costs is independent of the contestant's own effective size. Furthermore, it assumes that the rate of incurring costs from the rival is proportional to the initial effective size ratio---such that the larger contestant inflicts costs faster but incurs them slower. Importantly, the integration of costs into the model, using Eq. \eqref{eq14}, is not predicated on these simplifying assumptions. Combining Eqs. \eqref{eq14} and \eqref{eq15}, the resulting expressions for $m_i$ and $m_j$ during the contest are
\begin{equation} \label{eq16}
    m_i(t_\sigma) = \frac{\mu_i}{1 + \left(K_\mathrm{self} + K_{ij}\,\dfrac{\mu_j}{\mu_i}\right)t_\sigma},\qquad m_j(t_\sigma) = \frac{\mu_j}{1 + \left(K_\mathrm{self} + K_{ij}\,\dfrac{\mu_i}{\mu_j}\right)t_\sigma}.
\end{equation}
While both $m_i$ and $m_j$ decrease monotonically with $t_\sigma$, when $\mu_i > \mu_j$ and $K_{ij} \ne 0$ (e.g., in an asymmetric contest with mutual assessment), the ratio $m_i/m_j$ increases and approaches a constant as $t_\sigma \rightarrow \displaystyle\infty$ (Supporting  Information S5). The dynamics of $m_i$, $m_j$, and $m_i/m_j$ according to Eq. \eqref{eq16} is shown with only self-inflicted costs ($K_{ij} = 0$) in Fig. \ref{fig4}\textit{A}, and with both self-inflicted and rival-inflicted costs ($K_{ij} \ne 0$) in Fig. \ref{fig4}\textit{B}.

The dependence of the effective sizes on $t_\sigma$ due to costs means that the interaction potentials themselves depend on $t_\sigma$ through Eq. \eqref{eq13}, as demonstrated in Fig. \ref{fig4}\textit{C,D}. This ultimate manifestation of costs in the model has a straight-forward behavioural interpretation: the accumulation of costs affects the internal motivation of the contestants to fight and/or their perception of the opponent, and therefore alters the nature of their interaction. In pure self-assessment, self-inflicted costs lead to simultaneous reduction in the effective attraction of both contestants towards their rival, as evident by the diminishing potential wells of Fig. \ref{fig4}\textit{C}. In mutual assessment, rival-inflicted costs increase the asymmetry of the interaction, as captured by the opposing trends of the potentials in Fig. \ref{fig4}\textit{D}. The relative contest potential vary accordingly with $t_\sigma$, as shown in Fig. \ref{fig4}\textit{E,F}. For both cases studied here---pure self-assessment (Eq. \eqref{eq10} with $s = 1$) and pure mutual assessment (Eq. \eqref{eq11} with $s_Q = 1$ and $r_Q = 2$)---the bounding well of $V_\mathrm{contest}$, which defines the contest regime, 
becomes less attractive (shallower) as the contest proceeds. In Fig. \ref{fig4}\textit{E} this trend eventually leads to contest termination, as the bounding well disappears and $V_\mathrm{contest}$ becomes strictly repulsive.

\section*{\uppercase{Trends of Contest Duration}}
The duration of contests is widely used as a readily observable and relatively unbiased measure of contest dynamics. Many behavioural studies examine the relation between contest duration and the RHP of the contestants in order to determine whether a given species tends to settle contests through self- or mutual assessment \cite{chapin2019further,arnott2009assessment,pinto2019all}. Notably, pure self-assessment models predict that contest duration would generally increase with the mean RHP of the contestants \cite{taylor2003mismeasure}, as the persistence of purely self-assessing contestants is determined by their absolute RHP. In contrast, mutual assessment models predict contest duration to strongly decrease with the contestants' RHP ratio, and that the duration of RHP-matched contests would not scale with RHP at all \cite{taylor2003mismeasure}. Here we show that the same trends can be derived and understood within our model using a physics-inspired reasoning of 'contestant particles' escaping a potential well. 

When deriving these trends, we consider two limiting cases in terms of fighting costs. In the 'no-cost' limit, the costs accumulated during the course of a single contest are negligible. This means that the effective sizes $m_i$ and $m_j$ do not change during the contest, making $V_\mathrm{contest}$ effectively independent of contest duration. Although not realistic, the no-cost limit sets a useful upper bound on contest duration. In the opposing 'cost-driven' limit, a single contest entails very significant costs, resulting in $m_i$ and $m_j$ decaying substantially during the contest, and in a corresponding dependence of $V_\mathrm{contest}$ on contest duration---as illustrated in Fig. \ref{fig3}\textit{E},\textit{F}. The termination of a no-cost contest is noise-driven---entirely governed by the stochastic component of the contestants' motion and the shape of $V_\mathrm{contest}$, while the duration of a cost-driven contest is dominated by the accrued costs, as described below.

\begin{figure}[h!]
    \centering
    \includegraphics[width = 13.99cm]{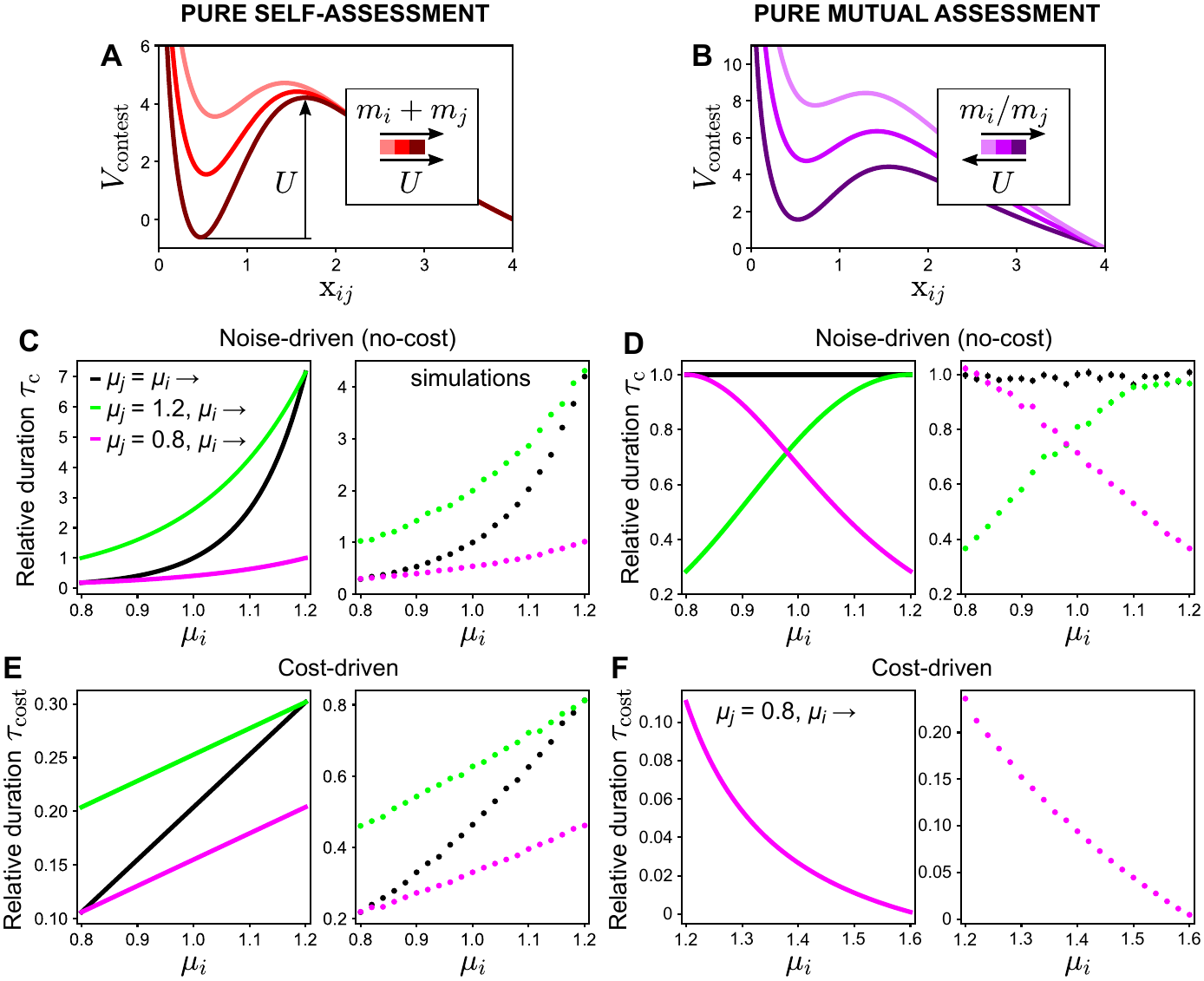}
    \caption{\textbf{Trends of contest duration.} (\textit{\textbf{A}},\textit{\textbf{B}}) The dependence of the effective 'contest bounding energy' $U$ on $m_i$ and $m_j$ in pure self- and pure mutual assessment. (\textit{\textbf{C}}--\textit{\textbf{F}}) Three trends of contest duration with respect to the initial effective sizes $\mu_i$ and $\mu_j$ are examined in the model for pure self- and pure mutual assessment: (1) Increasing size in size-matched contests ($\mu_j = \mu_i\rightarrow$), (2) Increasing size of smaller contestant in asymmetric contests ($\mu_j = 1.2,\,\mu_i\rightarrow$), and (3) Increasing size of larger contestant in asymmetric contests ($\mu_j = 0.8,\,\mu_i\rightarrow$). The \textit{Left} graph of each panel shows the qualitative theoretical trends calculated according to Eqs. \eqref{eq17}--\eqref{eq19} with $\Pi = \Pi^\star = 1$ and $T_\mathrm{eff} = 2D = 1$ (see also Supporting Information S6 and S8), and the \textit{Right} graph shows the mean trends in simulations ($n = 5\,000$, error bars show SEM). (\textit{\textbf{C}},\textit{\textbf{D}}) Noise-driven contest termination, where $m_i$ and $m_j$ remain the same during the contest, and therefore $V_\mathrm{contest}$ is independent of contest duration. (\textit{\textbf{E}},\textit{\textbf{F}}) Cost-driven contest termination, where the decay of $m_i$ and $m_j$ during the contest is governed by Eq. \eqref{eq16}. The cost rates are $K_\mathrm{self} = 0.2$, $K_{ij} = 0$ in \textit{E}, and $K_\mathrm{self} = 0.01$, $K_{ij} = 0.3$ in \textit{F}. In \textit{F}, $\mu_i$ is varied in the range where the analytic expression for the cost-driven duration is valid (Supporting Information S8), given that $\mu_j = 0.8$.
    }\label{fig5}
\end{figure}

In the context of the model, contest duration can be thought of as the time it takes for the contestants to escape the potential well of $V_\mathrm{contest}$ once they are trapped by its transient bounded state. The effective 'contest bounding energy', $U$, can be obtained for any mode of assessment (Supporting Information S6) as a function of $m_i$ and $m_j$,
\begin{equation}
    \label{eq17}
    U(m_i, m_j) = \left[V_\mathrm{contest}(\mathrm{x_\cap}) - V_\mathrm{contest}(\mathrm{x_\cup})\right]_{m_i,\,m_j}
\end{equation}
where note that $V_\mathrm{contest}$, as well as $\mathrm{x_\cap}$ and $\mathrm{x_\cup}$, depend on $m_i$ and $m_j$ through Eq. \eqref{eq9}. Fig. \ref{fig5}\textit{A},\textit{B} illustrate the dependence of $U$ on $m_i$ and $m_j$ in pure self- and pure mutual assessment. Notably, in pure self-assessment $U$ increases with the sum of the effective sizes (for $s \ne 1$ in Eq. \eqref{eq10}, $U$ increases with the power sum ${m_i}^s + {m_j}^s$), whereas in pure mutual assessment, $U$ depends only on the effective size ratio, and decreases as $m_i/m_j$ increases (for proper values of $s_Q$ and $r_Q$ in Eq. \eqref{eq11}, e.g. for the chosen values of $s_Q = 1$ and $r_Q = 2$. See Supporting Information S7).

In the no-cost case, contest dynamics are analogous to the dynamics of Brownian particles in an energy trap of constant depth $U$ \cite{hanggi1990reaction}. Then the mean contest duration, $t_\mathrm{c}$, should increase exponentially with $U$,
\begin{equation}
    \label{eq18}
    t_\mathrm{c} = \Pi\exp\left(\frac{U(m_i, m_j)}{T_\mathrm{eff}}\right)
\end{equation}
where the pre-exponential factor $\Pi$ is assumed to vary weakly with $m_i$ and $m_j$---such that the exponential dictates the qualitative trend of $t_\mathrm{c}$, and $T_\mathrm{eff}$ is the 'effective temperature' of the contestant particles' interaction. Eq. \eqref{eq18} relies on the assumption that, once within the contest range, the contestants reach the minimum of $V_\mathrm{contest}$ quickly---within an average time that is significantly shorter than $t_\mathrm{c}$---and spend the rest of the contest near the minimum until they escape, as in Fig. \ref{fig2}\textit{B}.

In the cost-driven case, the cost-dominated contest duration, denoted $t_\mathrm{cost}$ henceforth, can be obtained by considering the decay of the effective sizes due to costs up to the point where $V_\mathrm{contest}$ is no longer attractive, as illustrated in Fig. \ref{fig4}\textit{E}. This happens when the two extrema of $V_\mathrm{contest}$ merge into a single inflection point, which for our particular choice of interaction potentials satisfies the condition (Supporting Information S1)
\begin{equation}\label{eq19}
    \frac{\alpha_{j\rightarrow i} + \alpha_{i\rightarrow j}}{\delta_{j\rightarrow i} + \delta_{i\rightarrow j}} = \frac{e}{2},
\end{equation}
where $e$ is Euler's number. Together with Eqs. \eqref{eq9} and \eqref{eq16}, the condition of Eq. \eqref{eq19} can be invoked to evaluate $t_\mathrm{cost}$ for any mode of assessment. In Supporting Information S8, we derive analytic expressions for $t_\mathrm{cost}$ under pure self-assessment and pure mutual assessment, which are valid estimates when the costs are large enough to make $t_\mathrm{cost}$ much smaller than the noise-driven duration $t_\mathrm{c}$.

Since we are interested in the trends of contest duration with respect to $m_i$ and $m_j$ under a particular assessment strategy, and since absolute durations vary between species, it is useful to define relative durations as $\tau_\mathrm{c} = t_\mathrm{c}/t_\mathrm{c}^\star$ and $\tau_\mathrm{cost} = t_\mathrm{cost}/t_\mathrm{c}^\star$, with $t_\mathrm{c}^\star = {\Pi^\star} \exp(U^\star/T_\mathrm{eff})$, where $\Pi^\star$ and $U^\star$ correspond to a no-cost contest between two size-matched contestants of the reference effective size ($m_i = m_j = 1$, recall Eq. \eqref{eq8}), for which all modes of assessment described by Eq. \eqref{eq9} yield identical interaction potentials, and therefore the same mean contest duration $t^\star_\mathrm{c}$.

We examine three trends of contest duration with respect to the initial effective sizes, $\mu_i$ and $\mu_j$, for pure self- and pure mutual assessment: (1) Increasing size in size-matched contests, (2) Increasing size of the smaller contestant in asymmetric contests, and (3) Increasing size of the larger contestant in asymmetric contests, as described in Fig. \ref{fig5}\textit{C}--\textit{F}. The theoretical trends predicted by Eqs. \eqref{eq17}--\eqref{eq19} are in agreement with the mean trends obtained from simulations, and are aligned with previously reported experimental trends and predictions of game-theoretic models \cite{arnott2009assessment,chapin2019further}. This demonstrates that contest duration can be derived directly from the underlying physical properties of the interaction, as an emergent feature of contest dynamics. Comparison between the no-cost trends (Fig. \ref{fig5}\textit{C},\textit{D}) and the cost-driven trends (Fig. \ref{fig5}\textit{E},\textit{F}) shows that both cases give rise to the same qualitative trends---suggesting that explicit integration of fighting costs (which relies on a cost function that typically cannot be measured) is not necessary to explain these previously observed trends. 

\section*{\uppercase{Properties of asymmetric contests}}

\subsection*{Dynamics of asymmetric contests}
\begin{figure}[h!]
    \centering 
    \includegraphics[width = 14.1cm]{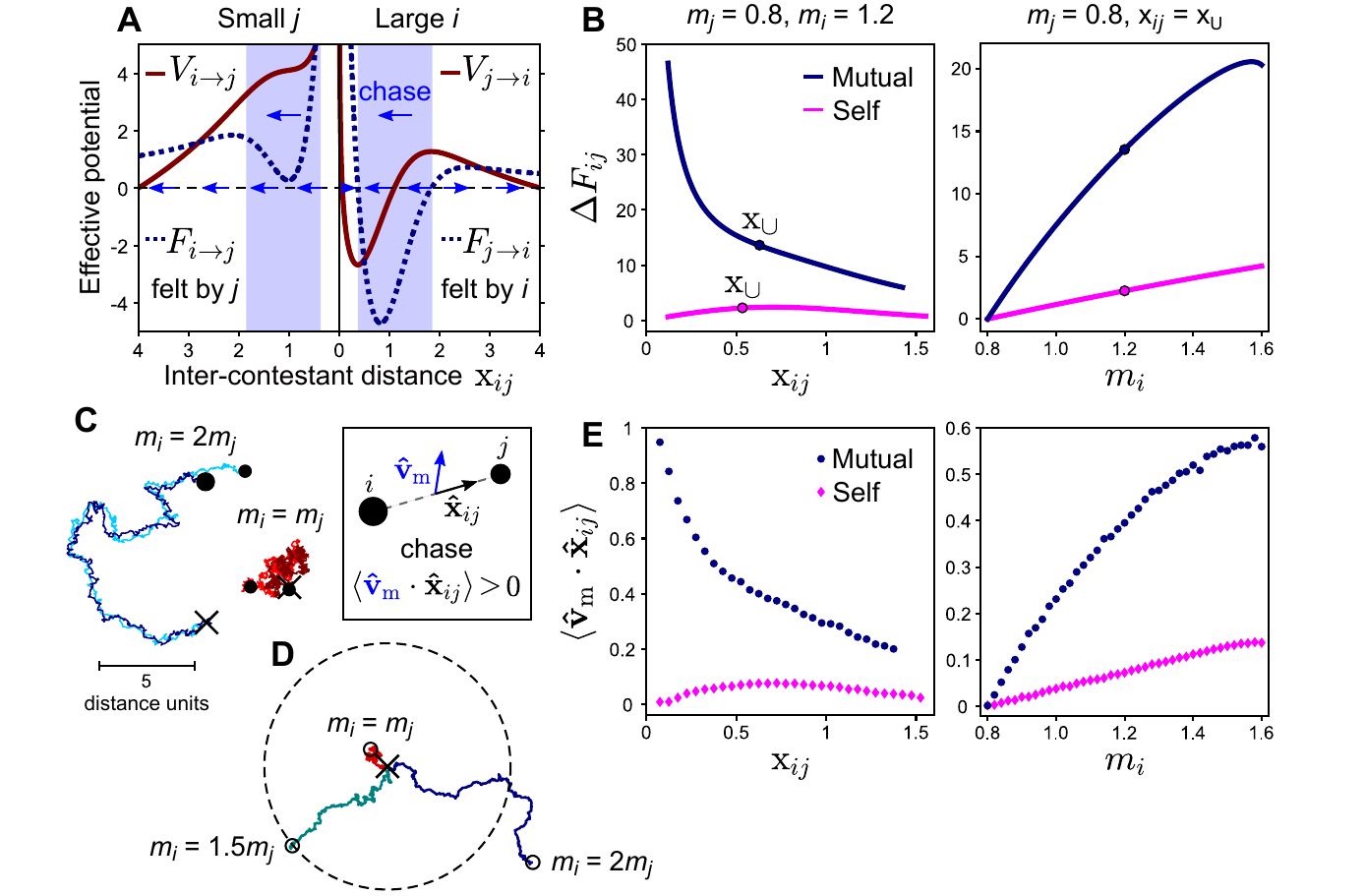}
    \caption{\textbf{Interaction asymmetry and chase dynamics.} (\textit{\textbf{A}}) Strongly asymmetric interaction potentials illustrate how interaction asymmetry leads to chase dynamics. Shaded rectangles mark the chase range, within which the larger contestant is attracted by a deep minimum, while the smaller contestant is repelled in the same direction. Arrows indicate the direction of motion for each contestant, as dictated by the signs of the forces $F_{j\rightarrow i}$ and $F_{i\rightarrow j}$. (\textit{\textbf{B}}) Force asymmetry $\Delta F_{ij} = F_{i\rightarrow j} - F_{j\rightarrow i}$ as a function of $\mathrm{x}_{ij}$ for strongly unmatched contestants (\textit{Left}) and $\Delta F_{ij}(\mathrm{x}_\cup)$ (at the minimum of $V_\mathrm{contest}$) as a function of the larger contestant's effective size $m_i$ (\textit{Right}) under pure self-assessment (Eq. \eqref{eq10} with $s = 1$) and pure mutual assessment (Eq. \eqref{eq11} with $s_Q = 1$ and $r_Q = 2$). Circles mark equivalent points in both graphs. (\textit{\textbf{C}}) Typical trajectories of strongly asymmetric and symmetric contests in simulations. Chase dynamics is quantified by the temporal mean of the direction correlation between $\hat{\mathbf{v}}_\mathrm{m}$ and $\hat{\mathbf{x}}_{ij}$, as defined in the inset. (\textit{\textbf{D}}) Trajectories of the contestants' midpoint during contests under pure mutual assessment for different effective size ratios. These trajectories, of equal durations, demonstrate the effect of chase dynamics on the contestants' displacement during the contest. In \textit{C} and \textit{D}, the 'X' marks the contestants' midpoint at the contest onset and the circles mark final positions. (\textit{\textbf{E}}) The 'chase correlator' $\langle \hat{\mathbf{v}}_\mathrm{m}\cdot \hat{\mathbf{x}}_{ij} \rangle$ as calculated from simulations ($n = 5\,000$) under pure self-assessment and pure mutual assessment, where $\mathrm{x}_{ij}$ and $m_i$ are varied as in \textit{B}.
    }
    \label{fig6}
\end{figure}

Interactions between unmatched contestants are characterized by broken symmetry. Here we explore in the model the effects of this asymmetry on the spatio-temporal dynamics of the contest and the emergence of chase dynamics---a ubiquitous feature of animal contests. As illustrated in Fig. \ref{fig6}\textit{A} for strongly asymmetric interaction potentials, within the 'chase' range the larger contestant is attracted by a deep minimum---leading it to move towards its smaller rival, while the smaller contestant is repelled in the same direction---moving away from its larger rival. Together, these effects gives rise to directed chase behaviour during the contest, where the larger contestant tends to chase its smaller rival away. The extent to which a contest is dominated by chase dynamics is determined by the magnitude of asymmetry between the inter-contestant effective forces, measured by
\begin{equation}\label{eq20}
    \Delta F_{ij}(\mathrm{x}_{ij}) = 
    F_{i\rightarrow j}(\mathrm{x}_{ij}) - F_{j\rightarrow i}(\mathrm{x}_{ij}).
\end{equation}

In particular, since contests are governed by the forces near the minimum of $V_\mathrm{contest}$, the extent of chase dynamics during a contest is determined by $\Delta F_{ij}(\mathrm{x}_\cup)$. Fig. \ref{fig6}\textit{B} shows $\Delta F_{ij}$ as a function of $\mathrm{x}_{ij}$ for strongly unmatched contestants (\textit{Left}) and $\Delta F_{ij}(\mathrm{x}_\cup)$ as a function of the larger contestant's effective size $m_i$ (\textit{Right}) under pure self-assessment (Eq. \eqref{eq10} with $s = 1$) and pure mutual assessment (Eq. \eqref{eq11} with $s_Q = 1$ and $r_Q = 2$). This comparison further illustrates that in mutual assessment, a relatively small effective size difference is translated into strong interaction asymmetry.

To quantify chase dynamics in the contestants' trajectories, we consider the direction correlation between the velocity of the contestants' midpoint---measured by $\hat{\mathbf{v}}_\mathrm{m}$, and the inter-contestant direction $\hat{\mathbf{x}}_{ij}$, as defined in Fig. \ref{fig6}\textit{C} (\textit{Inset}). The temporal mean of this 'chase correlator', denoted by $\langle \hat{\mathbf{v}}_\mathrm{m}\cdot \hat{\mathbf{x}}_{ij} \rangle$, is theoretically zero for symmetric interactions, but positive for asymmetric interactions because of the directional bias imposed by the asymmetry, and approaches 1 if the interaction is dominated by the chase phase. 

Fig. \ref{fig6}\textit{C},\textit{D} compare typical trajectories of strongly asymmetric and symmetric contests (under pure mutual assessment) in simulations. While the trajectories of symmetric contests are scrambled and relatively localized, the trajectories of strongly asymmetric contests feature substantial directional alignment and persistence, and consequently much greater displacement, due to chase dynamics. Fig. \ref{fig6}\textit{E} shows the chase correlator $\langle \hat{\mathbf{v}}_\mathrm{m}\cdot \hat{\mathbf{x}}_{ij} \rangle$ as calculated from simulations, where $\mathrm{x}_{ij}$ and $m_i$ are varied as in Fig. \ref{fig6}\textit{B}. Evidently, the trends of $\langle \hat{\mathbf{v}}_\mathrm{m}\cdot \hat{\mathbf{x}}_{ij} \rangle$ mirror the trends of $\Delta F_{ij}$, suggesting that the relation between $\langle \hat{\mathbf{v}}_\mathrm{m}\cdot \hat{\mathbf{r}}_{ij} \rangle$ and $\Delta F_{ij}$ follows simple (linear) scaling within almost the entire contest range and for a wide range of asymmetries.

\vspace{-10pt}

\subsection*{Contest outcome and size-related advantage}
\begin{figure}[h!]
    \centering 
    \includegraphics[width = 14.1cm]{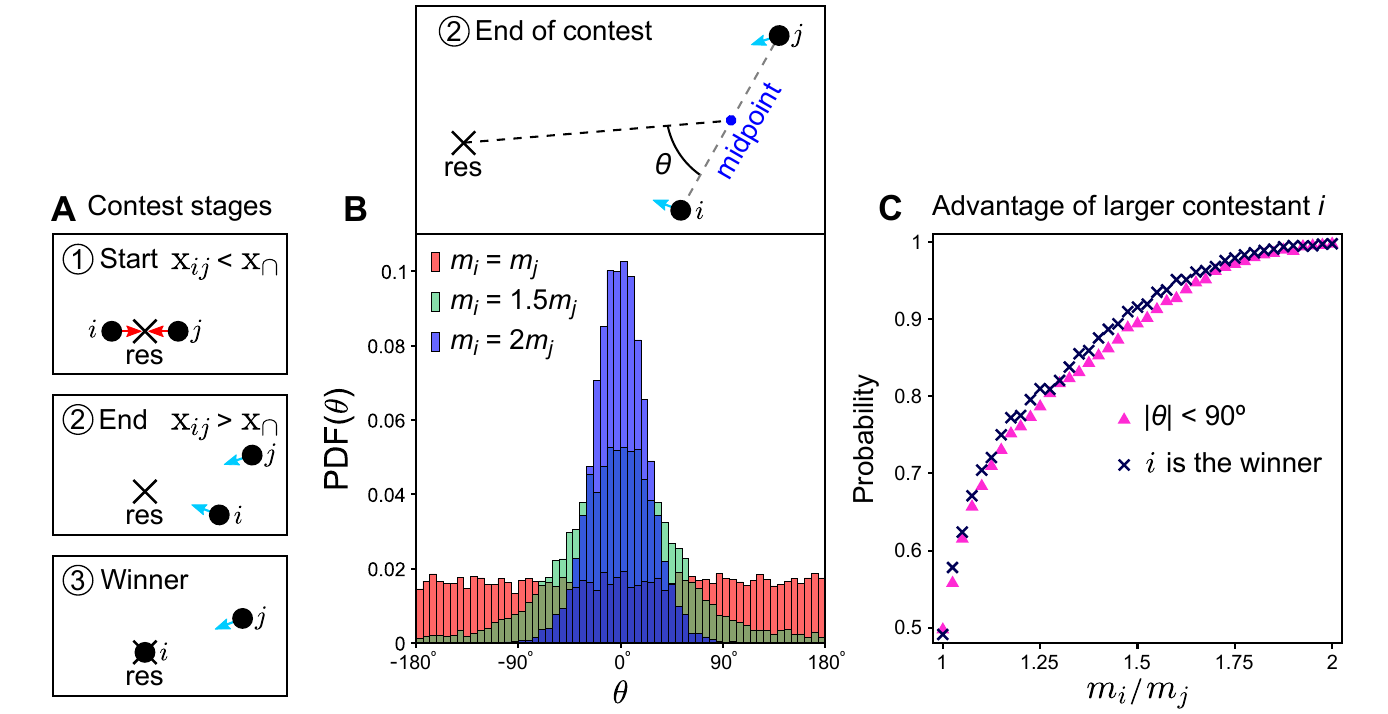}
    \caption{\textbf{Winning the contest.} (\textit{\textbf{A}}) The stages of a 'fair start' contest, which starts with the contestants at equivalent vantage points with respect to the resource---as in \textcircled{1}, ends at some configuration---as in \textcircled{2}, and subsequently its outcome (winner) is determined when one of the contestants reaches the resource---as in \textcircled{3}. (\textit{\textbf{B}}) The positional advantage of contestant $i$ with respect to the resource at the end of the contest can be quantified by the angle $\theta$, as defined in the upper panel. When $|\theta|>90^{\circ}$, $i$ is closer to the resource than $j$. The lower panel shows the distribution of $\theta$ under pure mutual assessment (Eq. \eqref{eq11} with $s_Q = 1$ and $r_Q = 2$) for different effective size ratios. The bias towards $\theta = 0^{\circ}$ increases with effective size asymmetry due to increasing chase dynamics, where $i$ chases $j$ away from the resource. (\textit{\textbf{C}}) Comparison between the probability that the larger contestant $i$ is closer to the resource at the end of the contest ($|\theta|<90^{\circ}$) and the probability that $i$ is the winner, as a function of the size ratio. In \textit{B} and \textit{C}, $n = 10\,000$ simulations for each effective size ratio.
    }
    \label{fig7}
\end{figure}

The ultimate expression of the RHPs is the outcome of the contest, wherein the winner obtains the resource. In the model, the resource is well-defined as the global minimum of the effective resource potential, and the outcome of the contest is determined when one of the contestants reaches the resource after the contest has ended. The advantage of larger effective size can be therefore examined directly in the model. In Fig. \ref{fig7}, we demonstrate for pure mutual assessment (Eq. \eqref{eq11} with $s_Q = 1$ and $r_Q = 2$) how interaction asymmetry increases the winning probability of the larger contestant---further showcasing the scope of the model in the describing observable contest dynamics. 

We consider 'fair start' contests (Fig. \ref{fig7}\textit{A}), which start when the contestants reach the contest range at equivalent vantage points with respect to the resource---such that neither of them has an initial positional advantage. Any advantage at winning such contests would therefore emerge from the dynamics of the contest itself. In Fig. \ref{fig7}\textit{B}, we examine how the distribution of the contestants' positions at the end of the contest is affected by the size asymmetry. As the size asymmetry increases, contests increasingly end with the larger contestant at positional advantage with respect to the resource ($|\theta|<90^{\circ}$, see upper panel in Fig. \ref{fig7}\textit{B}). This effect is due to increasing chase dynamics, in which the larger contestant tends to chase its rival away from the resource---leaving the larger contestant closer to the resource when the contest terminates. Fig. \ref{fig7}\textit{C} compares the probability that the larger contestant is closer to the resource at the end of the contest with the probability that it is the winner, as a function of the size ratio. The match between these probabilities means that in the model's simulations, the positional configuration at the end of the contest determines its outcome, where the contestant closer to the resource becomes the winner.

\section*{DISCUSSION}
The study of animal contests has a rich history, extending back to the first formal application of evolutionary stable strategies to explain the evolutionary logic of these agonistic interactions \cite{smith1973logic,harman2011birth}. In the following decades, game-theoretic models have yielded great insights into animal contests across taxonomic boundaries and contexts \cite{milinski1991competition,briffa2013introduction}. Nevertheless, the scarcity of direct empirical evidence in support of these models highlights the limitations of game-theoretic approaches in the study of animal contests  \cite{taylor2003mismeasure,pinto2019all,chapin2019further}. In particular, the difficulty of measuring contest dynamics as modelled within game theory has been a topic of recent discussion and debate \cite{taylor2003mismeasure,chapin2019further,parker2019so,elwood2019problems,leimar2019game}. One inherent barrier in the correspondence between theory and experiment has been the omission of within-contest dynamics by game-theoretic models, which typically verify their underlying decision rules based on contest end-points.

In this work, we developed a new theoretical framework for animal contest dynamics based on measurable spatio-temporal characteristics of contest behaviour. The basic building blocks of our model are effective interaction potentials, which encode generic rules of inter-contestant motion through the effective interaction forces that they generate. Importantly, these rules of motion can be directly extracted from empirical measurements of contest trajectories, and the simulated contest dynamics that our model gives rise to can be directly compared with the observable dynamics of real contests. Our approach requires few \textit{a priori} assumptions about the observed interactions and is not predicated on decision rules imposed by a specific assessment strategy. Moreover, it has greater applicability to the study of contests in natural conditions because it can utilize the entire information provided by the contestants' trajectories, and does not critically rely on the ability to measure contest duration. Our approach is therefore robust to partially sampled contests, in which the observer may be unsure of the onset of the contest, where the contestants may move out of sight, or when contests are prematurely ended by an external stimulus.

Once the basic form of the effective interaction potentials has been established, our framing of assessment strategies as variations of the interaction parameters depending on the contestants’ RHPs, as represented by the assessment function, allows the exploration of various strategies along the assessment continuum. This includes strategies that do not fall into the self- or mutual assessment categories, such as mixed assessment strategies \cite{chapin2019further}. Our model's representation of assessment strategies also brings them into the measurable domain where, rather than being retrospectively assigned based on the eventual contest outcome, the assessment strategy can in principle be inferred from the way the interaction varies with the contestants' RHPs. More broadly, modeling approaches that relate the measurable nature of the interaction to the current state of the contestants can be extended to encompass other determinants of behaviour, as we have demonstrated in our implementation of fighting costs.

While our model offers a new way of deriving and understanding previously described strategy-dependent trends of contest duration as a function of the contestants' RHPs, the detailed description of the contestants' spatial dynamics allows us to go far beyond paraphrasing the predictions of other models. We showcased some of these applications by exploring spatio-temporal properties of asymmetric contests, notably the emergence of chase dynamics and their contribution to size-related advantage. Such applications are especially useful to the study of complex competitive scenarios, for example systems in which the spatial density of interacting contestants is highly variable \cite{haluts2021spatiotemporal}.

Furthermore, the ability to extract effective interaction potentials directly from contest trajectories offers a robust method for detecting changes within an interaction over time. This can be done by sampling the trajectories of the same contestants at different times---for example by dissecting the contests into their initial and terminal stages, or by separately analyzing consecutive contests among the same rivals in the same or in different conditions. Such data can be used to determine how strategies update over time or context, which has been previously discussed as a useful contribution to the understanding of assessment strategies \cite{chapin2019further}. More generally, by monitoring the detailed dynamics of contests, we can study the role that injuries, energetic expenditure, learning from past experience, or an intrinsic change in motivation play in the behavioural outputs of contestants, and thereby gain insights into the mechanisms that underlie these outputs \cite{reichert2017cognition}. The above suggests many further applications of our modeling approach, which we leave for future work.

With the rapid development of tools to gather extremely high-resolution data on behavioural interactions in both laboratory \cite{pereira2022sleap,mathis2018deeplabcut} and field settings \cite{francisco2020high,nathan2022big}, our model aims to bridge the growing gap between empirical capabilities and theory. Not only can the model recapitulate many predictions of previous models, but importantly it enables the description of realistic and detailed contest dynamics within observable time scales. Our theoretical framework should therefore facilitate new approaches and insights in the study of this widespread aspect of animal behaviour.
\bibliography{sample}

\clearpage
\vspace{-25pt}
\section*{SUPPORTING INFORMATION}

\subsection*{S1. Extrema of the interaction potential}
The effective potential of main text Eq. \eqref{eq1} has the form 
\begin{equation}\label{eqS1}\tag{S1}
    V(\mathrm{x}) = - \mathcal{A}\exp(-\mathcal{B}\mathrm{x}^2) - \mathcal{D}\ln(\mathrm{x})
\end{equation}
The derivative of Eq. \eqref{eqS1} with respect to $\mathrm{x}$ is
\begin{equation}\label{eqS2}\tag{S2}
    \frac{dV(\mathrm{x})}{d\mathrm{x}} = 2\mathcal{A}\mathcal{B}\,\mathrm{x}\exp(-\mathcal{B}\mathrm{x}^2) - \frac{\mathcal{D}}{\mathrm{x}}
\end{equation}
Equating Eq. \eqref{eqS2} to zero and rearranging, we get
\begin{equation}\label{eqS3}\tag{S3}
    -\mathcal{B}\mathrm{x}^2\exp(-\mathcal{B}\mathrm{x}^2) = -\frac{\mathcal{D}}{2\mathcal{A}}
\end{equation}
Eq. \eqref{eq3} has the form $\omega\exp(\omega) = z$, which holds only if $\omega = W_k(z)$, where $W_k$ (for $k\in\mathbb{Z}$) are branches of the Lambert $W$ function. We get
\begin{equation}\label{eqS4}\tag{S4}
    -\mathcal{B}\mathrm{x}^2 = W_k\left(-\frac{\mathcal{D}}{2\mathcal{A}}\right)\quad\Rightarrow\quad\mathrm{x} = \sqrt{-\frac{1}{\mathcal{B}}W_k\left(-\frac{\mathcal{D}}{2\mathcal{A}}\right)}
\end{equation}
In our case, the two real branches $W_{-1}$ and $W_0$ are the relevant solutions, and we have
\begin{equation}\label{eqS5}\tag{S5}
     \mathrm{x}_\cap = \sqrt{-\frac{1}{\mathcal{B}}W_{-1}\left(-\frac{\mathcal{D}}{2\mathcal{A}}\right)},\qquad
     \mathrm{x}_\cup = \sqrt{-\frac{1}{\mathcal{B}}W_0\left(-\frac{\mathcal{D}}{2\mathcal{A}}\right)}
\end{equation}
where $\mathrm{x}_\cap$ is a maximum and $\mathrm{x}_\cup$ is a minimum. These extrema exist if
\begin{equation}\label{eqS6}\tag{S6}
    \frac{\mathcal{A}}{\mathcal{D}} > \frac{e}{2}
\end{equation}
where $e$ is Euler's number. For $\mathcal{A}/\mathcal{D} = e/2$, the two extrema merge into a single inflection point, and for $\mathcal{A}/\mathcal{D} < e/2$, the potential becomes repulsive for all $\mathrm{x}$. With $\mathcal{A} = \alpha_{j\rightarrow i} + \alpha_{i\rightarrow j}$, $\mathcal{B} = \beta$, and $\mathcal{D} = \delta_{j\rightarrow i} + \delta_{i\rightarrow j}$, as in main text Eq. \eqref{eq5}, Eq. \eqref{eqS5} becomes
\begin{equation}\label{eqS7}\tag{S7}
     \mathrm{x}_\cap = \sqrt{-\frac{1}{\beta}W_{-1}\left(-\frac{\delta_{j\rightarrow i} + \delta_{i\rightarrow j}}{2\left(\alpha_{j\rightarrow i} + \alpha_{i\rightarrow j}\right)}\right)},\qquad
     \mathrm{x}_\cup = \sqrt{-\frac{1}{\beta}W_0\left(-\frac{\delta_{j\rightarrow i} + \delta_{i\rightarrow j}}{2\left(\alpha_{j\rightarrow i} + \alpha_{i\rightarrow j}\right)}\right)}
\end{equation}
which are the extrema of the contest potential of main text Eq. \eqref{eq5}.

\subsection*{S2. Inclusion of other (system-specific) features in $V_{j\rightarrow i}$}
\begin{figure}[h!]
    \renewcommand{\thefigure}{S1}
    \centering 
    \includegraphics[width = 14.1cm]{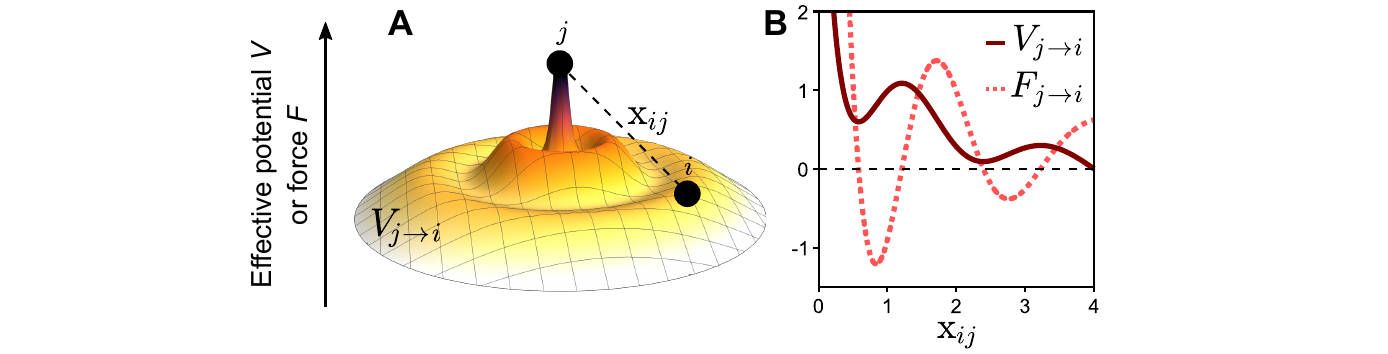}
    \caption{\textbf{Contestant interaction potential with a two-stage escalation.} (\textit{\textbf{A}}) The landscape of the effective 'contestant interaction potential' $V_{j\rightarrow i}$, as given in Eq. \eqref{eqS8}, with $\alpha_{j\rightarrow i} = 7$, $\delta_{j\rightarrow i} = 3$, $\beta = 1$, $\alpha_{\mathrm{eval}\,j\rightarrow i} = 1.7$, $\beta_\mathrm{eval} = 1$, and $\mathrm{x}_\mathrm{eval} = 2$. (\textbf{\textit{B}}) The profiles of $V_{j\rightarrow i}$ and its corresponding force $F_{j\rightarrow i} = -d V_{j\rightarrow i}/d\mathrm{x}_{ij}$ as a function of the distance between the contestants $\mathrm{x}_{ij}$. The graph of $V_{j\rightarrow i}$ was shifted vertically such that the lowest shown point has a 'height' of zero. Parameter values as in \textit{A}.
    }
    \label{figS1}
\end{figure}

The effective potential of main text Eq. \eqref{eq1} encodes minimal generic features of contest behaviour, but in some cases it may be important to include more details in order to fit the dynamics of real contests. To demonstrate the inclusion of additional features in $V_{j\rightarrow i}$, we consider contests that escalate through an intermediate 'evaluation' stage that precedes the ultimate escalation into a short-range interaction, as in ref. \cite{haluts2021spatiotemporal}. This feature can be added to $V_{j\rightarrow i}$ as an additional attractive Gaussian well centered at some $\mathrm{x}_\mathrm{eval} > 0$, yielding a contestant interaction potential of the form
\begin{equation}\label{eqS8}\tag{S8}
    V_{j\rightarrow i}(\mathrm{x}_{ij}) = - \alpha_{j\rightarrow i}\exp(-\beta{\mathrm{x}_{ij}}^2) - \alpha_{\mathrm{eval}\,j\rightarrow i}\exp(-\beta_\mathrm{eval}(\mathrm{x}_{ij} - \mathrm{x}_\mathrm{eval})^2) - \delta_{j\rightarrow i}\ln(\mathrm{x}_{ij})
\end{equation}
where the parameters $\alpha_{\mathrm{eval}\,j\rightarrow i}$, $\beta_\mathrm{eval}$, and $\mathrm{x}_\mathrm{eval}$ define the intermediate attractive well. With an appropriate choice of parameters, Eq. \eqref{eqS8} encodes a two-stage escalation, as illustrated in Fig. \ref{figS1}.

\vspace{-18pt}
\subsection*{S3. Simulations of contestant dynamics}
In two-dimensions, the spatio-temporal dynamics of a 'contestant particle' $i$ is governed by the following (overdamped) Langevin equation
\begin{equation}\label{eqS9}\tag{S9}
    \frac{d}{dt}\mathbf{r}_i = \frac{d}{dt}\begin{pmatrix}x_i\\y_i\end{pmatrix} = -\eta\begin{pmatrix}d/d x_i\\d/d y_i\end{pmatrix}V_{\mathrm{tot}\rightarrow i} + \sqrt{2D}\begin{pmatrix}\xi_{x_i}\\\xi_{y_i}\end{pmatrix}
\end{equation}

On the right-hand side of Eq. \eqref{eqS9}, the first term accounts for the deterministic driving force felt by contestant $i$ at position $\mathbf{r}_i = (x_i, y_i)$ due to $V_{\mathrm{tot}\rightarrow i}$ of main text Eq. \eqref{eq2}, where $\eta$ is the contestant's 'mobility'---which sets its response to the effective forces encoded by $V_{\mathrm{tot}\rightarrow i}$. The second term is the stochastic component of the translational motion, where $D$ is an effective diffusion coefficient, and $\xi_{x_i}$, $\xi_{y_i}$ are mutually uncorrelated sources of standard Gaussian white noise for each of the spatial dimensions. In the above description, we treat our contestants as 'passive' Brownian particles by assuming that all of the deterministic activity of a contestant is sufficiently encoded into $V_{\mathrm{tot}\rightarrow i}$. An 'active' component of the motion, notably directional persistence that is uncorrelated with the effective interaction forces, can be added to the dynamics of Eq. \eqref{eqS9} in order to match the characteristics of motion of particular animal contestants \cite{haluts2021spatiotemporal}. Note that the two-dimensional treatment is appropriate for contests taking place in an approximately planar environment. For contests characterized by substantial three-dimensional motion, such as aerial bird fights, Eq. \eqref{eqS9} can be trivially extended to three-dimensions, with $\mathbf{r}_i = (x_i, y_i, z_i)$, by considering a three-dimensional version of $V_{\mathrm{tot}\rightarrow i}$.

We used the two-dimensional dynamics of Eq. \eqref{eqS9}, with $\eta = 1$ and $D = 0.5$, to simulate all contest interactions described in this work. 
Importantly, $\eta$ and $D$ were set such that simulated contestants get in and out of the contest regime (main text Fig. \ref{fig1}\textit{F}) within a relatively short time---as in typical animal contests.

The first-order (Euler) discrete-time approximation for the time evolution of Eq. \eqref{eqS9} is given by

\vspace{-15pt}
\begin{equation}\label{eqS10}\tag{S10}
     \begin{pmatrix}x_i(t + \Delta t)\\y_i(t + \Delta t)\end{pmatrix} = \begin{pmatrix}x_i(t)\\y_i(t)\end{pmatrix} + \eta\begin{pmatrix}F_{x_i}(x_i(t))\\F_{y_i}(y_i(t))\end{pmatrix}\Delta t + \sqrt{2D\Delta t}\begin{pmatrix}\xi_{x_i}(t)\\\xi_{y_i}(t)\end{pmatrix}
\end{equation}
where $\Delta t$ is a sufficiently small, constant time step, such that $t = n\Delta t$ corresponds to the $n$-th step of the discrete-time sequence, and $F_{x_i} = -d V_{\mathrm{tot}\rightarrow i}/d x_i$. At every time step, the value of each noise source is independently drawn from a standard Gaussian distribution, where $\xi_{x_i}$, $\xi_{y_i} \sim \mathcal{N}(\mu = 0, \sigma^2 = 1)$. The parameter values $\eta = 1$ and $D = 0.5$, with a time step of $\Delta t = 0.005$, were used in Eq. \eqref{eqS10} for all the simulations in this work. The simulations were implemented in MATLAB R2019b. Note that the dynamics of contestant $j$ is governed by an equivalent equation for $x_j$ and $y_j$.

\subsection*{S4. Extracting interaction potentials from trajectories}
Given trajectories of two interacting contestants, we implement main text Eq. \eqref{eq6} to extract a relative velocity profile by binning and averaging the relative velocity along the trajectories according to the inter-contestant distance $\mathrm{x}_{ij}$,
\begin{equation} \label{eqS11}\tag{S11}
    v_\mathrm{rel}(\mathrm{x}_{ij}[k]) = \text{mean}\left(\frac{\Delta\mathrm{x}_{ij}}{\Delta t}(\mathrm{x}_{ij}[k])\right),\quad\text{bin }k = \left\{\mathrm{x}_{ij}\,\big|\,\mathrm{x}_k\leq\mathrm{x}_{ij}<\mathrm{x}_{k+1}\right\},\quad\mathrm{x}_{ij}[k] = \frac{\mathrm{x}_k + \mathrm{x}_{k+1}}{2}.
\end{equation}
We provide the following MATLAB R2019b code that implements Eq. \eqref{eqS11} to a set of contest trajectories. This code assumes that the data from each contest is stored as a MAT-file, where the contestant trajectories 'Trj1' and 'Trj2' are 2D arrays of length $N_\mathrm{frames}$. Plotting the output of this code yields a relative velocity profile as in main text Fig. \ref{fig2}\textit{D}.

\nolinenumbers
\begin{lstlisting}
%% extract averaged relative velocity profile from a set of contest trajectories
% calculation parameters (chosen according to data characteristics)
vbin = 1;                   % frame bin for velocity calculation
dt = 0.005;                 % time units per frame (e.g. 1/fps)
xbin = 0.01;                % inter-contestant distance bin size
xmax = 4;                   % max inter-contestant distance to use
xedges = 0 : xbin : xmax;   % bin edges for distance binning
% prepare for execution
folder = 'contest_trajectories\';
variables = {'Trj1', 'Trj2'};
folder_content = dir(folder);
collect_mvr = NaN(length(folder_content) - 2, length(xedges) - 1);
% loop through the files
for file_idx = 3 : length(folder_content)
    file_name = folder_content(file_idx).name;    
    load(strcat(folder, file_name), variables{:});
    % relative velocity calculation
    X1 = Trj1(:, 1); Y1 = Trj1(:, 2); X2 = Trj2(:, 1); Y2 = Trj2(:, 2); % contestant positions
    x = sqrt((X1 - X2).^2 + (Y1 - Y2).^2);                              % distances
    t = 0 : dt : (length(x) - 1) * dt;                                  % times
    Dx = x(vbin + 1 : vbin : end) - x(1 : vbin : end - vbin);           % distance diffs
    Dt = t(vbin + 1 : vbin : end) - t(1 : vbin : end - vbin);           % time diffs
    vr = Dx./Dt';                                                       % relative velocity
    % bin according to inter-contestant distance
    [~, ~, loc] = histcounts(x(1 : vbin : end - vbin), xedges);
    % get rid of empty bins 
    vr(~loc) = []; loc(~loc) = []; 
    % average relative velocities in bins
    mvr = accumarray(loc(:), vr(:)) ./ accumarray(loc(:), 1);
    % add to collection
    collect_mvr(file_idx - 2, 1 : length(mvr)) = mvr;
end
% averaged relative velocity profile
mvr = mean(collect_mvr, 'omitnan');
mx = (xedges(1 : end - 1) + xedges(2 : end))/2;
% get rid of NaN values (required for integration)
mx(isnan(mvr)) = []; mvr(isnan(mvr)) = [];

%% plot relative velocity profile
figure
plot(mx, mvr, '-m', 'LineWidth', 3)
title('averaged relative velocity profile')
xlabel('inter-contestant distance')
ylabel('relative velocity')
ylim([-5, 7]) % chosen for the given example
pbaspect([1.4, 1, 1])
\end{lstlisting}
We proceed by implementing main text Eq. \eqref{eq7} to obtain an averaged relative interaction potential $V_\mathrm{rel}$ by cumulative integration over the discrete values of $v_\mathrm{rel}$, using the trapezoidal rule,
\begin{equation} \label{eqS12}\tag{S12}
    \eta V_\mathrm{rel}(\mathrm{x}_{ij}[N]) = -\sum_{k=1}^N\frac{v_\mathrm{rel}(\mathrm{x}_{ij}[k-1]) + v_\mathrm{rel}(\mathrm{x}_{ij}[k])}{2}\Delta\mathrm{x}_{ij}[k]\approx-\int_0^{\mathrm{x}_{ij}} v_\mathrm{rel}(\mathrm{x}^\prime_{ij})\,d\mathrm{x}^\prime_{ij}
\end{equation}
where $\Delta\mathrm{x}_{ij}[k] = \mathrm{x}_{ij}[k] - \mathrm{x}_{ij}[k-1]$. The following MATLAB R2019b code implements Eq. \eqref{eqS12} and plots the result, yielding an averaged relative interaction potential as in main text Fig. \ref{fig2}\textit{D}.

\nolinenumbers
\begin{lstlisting}
%% calculate averaged relative interaction potential
mVr = cumtrapz(mx, -mvr); % cumulative integration via the trapezoidal rule

%% plot relative interaction potential (shifted by mVr(end))
figure
plot(mx, mVr - mVr(end), '-b', 'LineWidth', 3)
title('averaged relative interaction potential')
xlabel('inter-contestant distance')
ylabel('relative interaction potential')
ylim([-5, 7]) % chosen for the given example
pbaspect([1.4, 1, 1])
\end{lstlisting}
Now we can fit a model for $V_\mathrm{contest}$ to $V_\mathrm{rel}$---in this case main text Eq. \eqref{eq5}---in a restricted range. The fit range is bounded by $\mathrm{x}_{ij} = 0$ and an upper bound which is taken as the distant-most value of $\mathrm{x}_{ij}$ where $v_\mathrm{rel}$ changes sign. The following MATLAB R2019b code performs the fit and plots the result, along with $V_\mathrm{rel}$, as in main text Fig. \ref{fig2}\textit{E}.

\nolinenumbers
\begin{lstlisting}
%% fit model potential
syms 'x' 'a' 'b' 'c' 'd';               % create symbolic variables
model = -a*exp(-b*x^2) -d*log(x) + c;   % model interaction potential
fitfunc = matlabFunction(model);        % fit function
% fit options (this can dramatically affect the result)
startpoints = [1, 1, 0, 1];         % [a, b, c, d]
lowerbounds = [0.1, 0.1, 0.1, 0.1]; % [a, b, c, d]
xfit_upperb = 1.5;                  % fit range upper bound
fitopts = fitoptions('Method', 'NonlinearLeastSquares', ...
                     'Start', startpoints, ...
                     'Lower', lowerbounds, ...
                     'Exclude', mx > xfit_upperb);
% perform the fit
fV = fit(mx', (mVr - mVr(end))', fitfunc, fitopts);

%% plot relative interaction potential and fitted contest potential
figure
V_contest_fit = subs(model, [a, b, c, d], [fV.a, fV.b, fV.c, fV.d]);
V_shift = subs(V_contest_fit, x, max(mx)); % for shifting the potentials
plot(mx, mVr - mVr(end) - V_shift, '-b', 'LineWidth', 3)
hold on
fplot(matlabFunction(V_contest_fit - V_shift), [min(mx), max(mx)], '-r', 'LineWidth', 3)
title('fit to model')
xlabel('inter-contestant distance')
ylabel('interaction potential')
ylim([-5, 7]) % chosen for the given example
pbaspect([1.4, 1, 1])
\end{lstlisting}
The above code, along with a set of 30 simulated trajectories, is available in the following Zenodo repository:\\ \href{https://doi.org/10.5281/zenodo.7377238}{\textcolor{blue}{https://doi.org/10.5281/zenodo.7377238}}

\subsection*{S5. Dynamics of the effective sizes due to costs}
Recall main text Eq. \eqref{eq16},
\begin{equation} \label{eqS13}\tag{S13}
    m_i(t_\sigma) = \frac{\mu_i}{1 + \left(K_\mathrm{self} + K_{ij}\,\dfrac{\mu_j}{\mu_i}\right)t_\sigma},\qquad m_j(t_\sigma) = \frac{\mu_j}{1 + \left(K_\mathrm{self} + K_{ij}\, \dfrac{\mu_i}{\mu_j}\right)t_\sigma}
\end{equation}
The ratio between the effective sizes $m_i/m_j$ as a function of $t_\sigma$ is therefore given by
\begin{equation} \label{eqS14}\tag{S14}
    \frac{m_i}{m_j}(t_\sigma) = \frac{\mu_i}{\mu_j}\cdot\frac{1 + \left(K_\mathrm{self} + K_{ij}\, \dfrac{\mu_i}{\mu_j}\right)t_\sigma}{1 + \left(K_\mathrm{self} + K_{ij}\, \dfrac{\mu_j}{\mu_i}\right)t_\sigma}
\end{equation}
For $\mu_i > \mu_j$, $m_i/m_j$ increases with $t_\sigma$. The ratio approaches a constant $Q_{\displaystyle\infty}$ as $t_\sigma\rightarrow \displaystyle\infty$,
\begin{equation} \label{eqS15}\tag{S15}
    Q_{\displaystyle\infty} = \lim_{\displaystyle{t_\sigma\rightarrow\infty}}\left(\frac{m_i}{m_j}\right) = \frac{\mu_i}{\mu_j}\cdot\frac{K_\mathrm{self} + K_{ij}\, \dfrac{\mu_i}{\mu_j}}{K_\mathrm{self} + K_{ij}\, \dfrac{\mu_j}{\mu_i}}
\end{equation}

\subsection*{S6. Effective 'contest bounding energy'}
Combining main text Eqs. \eqref{eq5} and \eqref{eq19} with Eq. \eqref{eqS7}, and using the identity $\exp W(\mathrm{x}) = \mathrm{x}/W(\mathrm{x})$, the bounding potential well of the relative contest potential can be written as
\begin{equation}\label{eqS16}\tag{S16}
    U(m_i, m_j) = \left[V_\mathrm{contest}(\mathrm{x_\cap}) - V_\mathrm{contest}(\mathrm{x_\cup})\right]_{m_i,\,m_j} = \frac{\delta_{ij}}{2}\left(\ln\left(\dfrac{W_0\left(-\dfrac{\delta_{ij}}{2\alpha_{ij}}\right)}{W_{-1}\left(-\dfrac{\delta_{ij}}{2\alpha_{ij}}\right)}\right) + \dfrac{1}{W_{-1}\left(-\dfrac{\delta_{ij}}{2\alpha_{ij}}\right)} - \dfrac{1}{W_0\left(-\dfrac{\delta_{ij}}{2\alpha_{ij}}\right)}\right)_{m_i,\,m_j}
\end{equation}
where $\alpha_{ij} = \alpha_{j\rightarrow i} + \alpha_{i\rightarrow j}$ and $\delta_{ij} = \delta_{j\rightarrow i} + \delta_{i\rightarrow j}$. Note that $U(m_i, m_j)$ is independent of $\beta$. According to main text Eqs. \eqref{eq10} and \eqref{eq11}, we have in particular
\begin{equation}\label{eqS17}\tag{S17}
    \alpha_{ij} = \alpha_0\left({m_i}^s + {m_j}^s\right),\quad\delta_{ij} = 2\delta_0\quad\text{for pure self-assessment},
\end{equation}
and
\begin{equation}\label{eqS18}\tag{S18}
    \alpha_{ij} = \alpha_0\left(\left(\frac{m_i}{m_j}\right)^{s_Q} + \left(\frac{m_j}{m_i}\right)^{s_Q}\right),\quad\delta_{ij} = \delta_0\left(\left(\frac{m_i}{m_j}\right)^{r_Q} + \left(\frac{m_j}{m_i}\right)^{r_Q}\right)\quad\text{for pure mutual assessment}.
\end{equation}

\subsection*{S7. Proper values of $s_Q$ and $r_Q$ in pure mutual assessment}
\begin{figure}[h!]
    \renewcommand{\thefigure}{S2}
    \centering 
    \includegraphics[width = 14.1cm]{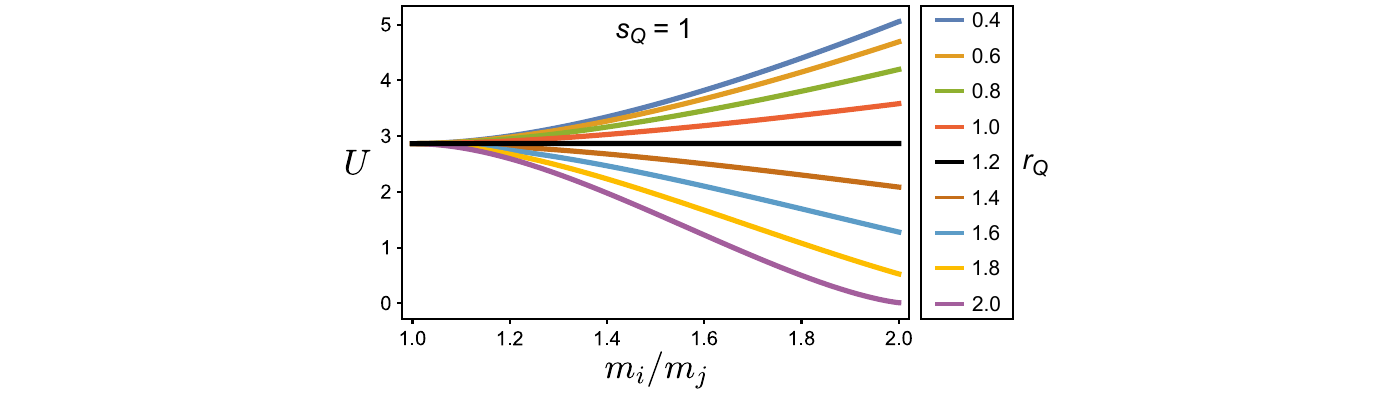}
    \caption{Contest bounding energy $U$ as a function of $m_i/m_j$, with $s_Q = 1$, $\alpha_0 = 7$, and $\delta_0 = 3$, for different values of $r_Q$.}
    \label{figS2}
\end{figure}

In pure mutual assessment, the contest bounding energy is expected to decrease monotonically with the size ratio, and this sets some restrictions on the values of the exponents $s_Q$ and $r_Q$. Clearly, for $U$ to decrease with $m_i/m_j$, the repulsive component of $V_\mathrm{contest}$ has to increase faster than its attractive component with $m_i/m_j$ (that is, in Eq. \eqref{eqS18}, $\delta_{ij}$ has to increase faster than $\alpha_{ij}$ with $m_i/m_j$). Since $\alpha_0 > \delta_0$ for $U$ to exist at all (see Eq. \eqref{eqS19} below), this means that $r_Q > s_Q$, but this is not strictly sufficient. Fig. \ref{figS2} shows $U$ as a function of $m_i/m_j$, with $s_Q = 1$, $\alpha_0 = 7$, and $\delta_0 = 3$, for different values of $r_Q$. With these parameter values, $U$ decreases monotonically with $m_i/m_j$ when $r_Q \gtrsim 1.2$.

\vspace{25pt}
\subsection*{S8. Cost-dominated contest duration}
When fighting costs are significant, the mean duration of a single contest is dominantly determined by the accumulated costs. A theoretical estimate for this cost-dominated duration (denoted henceforth as $t_\mathrm{cost}$) can be obtained by considering the decay of the effective sizes due to costs (Eq. \eqref{eqS13}) up to the point where the relative contest potential is no longer attractive. This happens when the extrema of Eq. \eqref{eqS7} merge into a single inflection point, that is when
\begin{equation}\label{eqS19}\tag{S19}
    \frac{\alpha_{j\rightarrow i} + \alpha_{i\rightarrow j}}{\delta_{j\rightarrow i} + \delta_{i\rightarrow j}} = \frac{e}{2}
\end{equation}

Under pure self-assessment, Eq. \eqref{eqS19} becomes
\begin{equation}\label{eqS20}\tag{S20}
    \frac{\alpha_0\left({m_i}^s + {m_j}^s\right)}{2\delta_0} = \frac{e}{2}
\end{equation}
Setting $K_{ij} = 0$ in Eq. \eqref{eqS13} (no rival-inflicted costs in pure self-assessment), we have
\begin{equation}\label{eqS21}\tag{S21}
    m_i(t_\sigma) = \frac{\mu_i}{1 + K_\mathrm{self}\,t_\sigma},\qquad m_j(t_\sigma) = \frac{\mu_j}{1 + K_\mathrm{self}\,t_\sigma}
\end{equation}
Combining Eqs. \eqref{eqS20} and \eqref{eqS21}, we get an expression for $t_\mathrm{cost}$ under pure self-assessment by solving for $t_\sigma$,
\begin{equation}\label{eqS22}\tag{S22}
    t_\mathrm{cost} = \frac{1}{K_\mathrm{self}}\left(\frac{\alpha_0}{e\delta_0}\left({\mu_i}^s + {\mu_j}^s\right)\right)^{1/s} - \frac{1}{K_\mathrm{self}}\quad\text{for pure self-assessment}.
\end{equation}
Note that for $s = 1$, $t_\mathrm{cost}$ increases linearly with the initial effective sizes.

Under pure mutual assessment, Eq. \eqref{eqS19} becomes
\begin{equation}\label{eqS23}\tag{S23}
    \frac{\alpha_0\left(\left(\dfrac{m_i}{m_j}\right)^{s_Q} + \left(\dfrac{m_j}{m_i}\right)^{s_Q}\right)}{\delta_0\left(\left(\dfrac{m_i}{m_j}\right)^{r_Q} + \left(\dfrac{m_j}{m_i}\right)^{r_Q}\right)} = \frac{e}{2}
\end{equation}
In general, Eq. \eqref{eqS23} can be solved numerically for $Q = m_i/m_j$, yielding a size ratio, $Q_0$, which satisfies Eq. \eqref{eqS19}. Given that $Q_0 < Q_\infty$, setting $Q(t_\sigma) = Q_0$ in Eq. \eqref{eqS14} and solving for the time yields $t_\mathrm{cost}$ under pure mutual assessment,
\begin{equation}\label{eqS24}\tag{S24}
    t_\mathrm{cost} = \frac{\dfrac{\mu_i}{\mu_j}\left(Q_0 - \dfrac{\mu_i}{\mu_j}\right)}{K_{ij}\left(\left(\dfrac{\mu_i}{\mu_j}\right)^3 - Q_0\right) - K_\mathrm{self}\,\dfrac{\mu_i}{\mu_j}\left(Q_0 - \dfrac{\mu_i}{\mu_j}\right)}\quad\text{for pure mutual assessment}.
\end{equation}
Note that Eq. \eqref{eqS24} is only valid if $Q_0 < Q_\infty$, and this requires a sufficiently large initial size ratio $\mu_i/\mu_j$. Otherwise, the size ratio $Q_0$ would never be reached during the contest, and thus the condition of Eq. \eqref{eqS19} cannot be satisfied.

With $s_Q = 1$ and $r_Q = 2$ in Eq. \eqref{eqS23}, $Q_0$ can be calculated analytically. Setting $Q = m_i/m_j$, Eq. \eqref{eqS23} with these power laws can be written as 
\begin{equation}\label{eqS25}\tag{S25}
    \left(Q + \frac{1}{Q}\right)^2 - \frac{2\alpha_0}{e\delta_0}\left(Q + \frac{1}{Q}\right) - 2 = 0
\end{equation}
Solving Eq. \eqref{eqS25} for $Q + 1/Q$, only one solution is positive,
\begin{equation}\label{eqS26}\tag{S26}
    Q + \frac{1}{Q} = \frac{\alpha_0}{e\delta_0} + \sqrt{\left(\frac{\alpha_0}{e\delta_0}\right)^2 + 2}
\end{equation}
which rearranges into a quadratic equation for $Q$,
\begin{equation}\label{eqS27}\tag{S27}
    Q^2 - \left(\frac{\alpha_0}{e\delta_0} + \sqrt{\left(\frac{\alpha_0}{e\delta_0}\right)^2 + 2}\right)Q + 1 = 0
\end{equation}
with the solutions
\begin{equation}\label{eqS28}\tag{S28}
    Q_\pm = \frac{b_0\pm\sqrt{{b_0}^2 - 4}}{2},\quad\text{where}\quad b_0 = \frac{\alpha_0}{e\delta_0} + \sqrt{\left(\frac{\alpha_0}{e\delta_0}\right)^2 + 2}
\end{equation}
Note that $Q_+ = 1/Q_-$, since
\begin{equation}\label{eqS29}\tag{S29}
    Q_+ = \frac{b_0 + \sqrt{{b_0}^2 - 4}}{2} = \dfrac{b_0 + \sqrt{{b_0}^2 - 4}}{2}\cdot\frac{b_0 - \sqrt{{b_0}^2 - 4}}{b_0 - \sqrt{{b_0}^2 - 4}} = \dfrac{4}{2\left(b_0 - \sqrt{{b_0}^2 - 4}\right)} = \frac{2}{b_0 - \sqrt{{b_0}^2 - 4}} = \frac{1}{Q_-}
\end{equation}
which means that if we set $m_i > m_j$, the effective size ratio corresponding to $Q_0$ is equal to $Q_+$, thus
\begin{equation}\label{eqS30}\tag{S30}
    Q_0 = \frac{b_0 + \sqrt{{b_0}^2 - 4}}{2}\quad\text{for mutual assessment with $s_Q = 1$ and $r_Q = 2$}.
\end{equation}
\end{document}